\documentclass[manuscript,screen]{acmart}
\AtBeginDocument{%
  }

\setcopyright{acmlicensed}
\copyrightyear{2018}
\acmYear{2018}
\acmDOI{XXXXXXX.XXXXXXX}
\acmConference[Conference acronym 'XX]{Make sure to enter the correct
  conference title from your rights confirmation email}{June 03--05,
  2018}{Woodstock, NY}
\acmISBN{978-1-4503-XXXX-X/2018/06}
\newcommand{\projname}{DGNA}

\usepackage{enumitem}
\usepackage{tcolorbox}
\usepackage{listings}
\definecolor{grey}{RGB}{128,128,128}
\definecolor{backcolor}{rgb}{0.99,0.99,0.99}
\newfloat{listing}{tbp}{lol}
\floatname{listing}{Listing}
\lstdefinestyle{gpucode}{%
    basicstyle=\footnotesize\ttfamily,
    numbers=left,
    breaklines=true,
    backgroundcolor=\color{backcolor},
    frame=none,
    xleftmargin=5pt,
    numbersep=1pt,
    columns=fullflexible,
    keepspaces=true,
    showstringspaces=false
}
\lstnewenvironment{minted}[2][]{\lstset{style=gpucode,language=#2,#1}}{}
\usepackage{subcaption}
\usepackage{multirow}
\usepackage{booktabs}
\newcommand{\tacorebuttal}[1]{\textcolor{black}{#1}}
\newcommand{\dramblackgpu}{922 cycles}
\renewcommand\footnotetextcopyrightpermission[1]{} 

\begin{document}

\title{DGNA: Dissecting GPU NUMA Architecture through Microbenchmarking and Data Analysis}

\author{Changxi Liu}
\email{changxi_liu@u.nus.edu}
\orcid{0000-0001-9240-5926}
\affiliation{%
  \institution{National University of Singapore}
  \city{Singapore}
  \state{}
  \country{Singapore}
}

\author{Yun Chen}
\email{yunchen@hkust-gz.edu.cn}
\orcid{0000-0001-8314-6985}
\affiliation{%
  \institution{The Hong Kong University of Science and Technology (Guangzhou)}
  \city{Guangzhou}
  \state{Guangdong}
  \country{China}
}

\author{Trevor E. Carlson}
\affiliation{%
  \institution{National University of Singapore}
  \country{Singapore}}
\email{tcarlson@nus.edu.sg}
\orcid{0000-0001-8742-134X}





\definecolor{rebuttalcolor}{RGB}{239,65,53}
\definecolor{darkgreen}{HTML}{466d1d}
\definecolor{orange}{HTML}{002654}
\definecolor{darkred}{RGB}{237,41,57}

\newcommand{\rebuttal}[1]{\textcolor{rebuttalcolor}{#1}}
\newcommand{\ignore}[1]{}
\newcommand{\dramamperegpuhigh}{546 cycles}
\newcommand{\dramamperegpulow}{385 cycles}
\newcommand{\dramamperegpugap}{161 cycles}

\newcommand{\dramhoppergpuhigh}{728 cycles}
\newcommand{\dramhoppergpuhighraw}{759 cycles}

\newcommand{\dramhoppergpulow}{554 cycles}

\newcommand{\dramhoppergpugap}{174 cycles}

\newcommand{\llcamperegpulow}{218 cycles}
\newcommand{\llcamperegpuhigh}{379 cycles}
\newcommand{\llcamperegpugap}{161 cycles}

\newcommand{\llchoppergpulow}{295 cycles}
\newcommand{\llchoppergpuhigh}{470 cycles}
\newcommand{\llchoppergpugap}{175 cycles}

\newcommand{\subllchoppergpulow}{280 cycles}
\newcommand{\subllchoppergpuhigh}{312 cycles}
\newcommand{\subllchoppergpugap}{32 cycles}

\definecolor{grey}{RGB}{239,65,53}
\begin{abstract}
Graphics Processing Units (GPUs), due to their immense parallel processing capabilities, have become essential across various fields, including gaming and artificial intelligence. With significant advancements in GPU cores, GPU memory efficiency has lagged, resulting in bottlenecks that can limit workload efficiency. To bridge this gap, a deep understanding of GPU memory architectures, particularly Non-Uniform Memory Access (NUMA) mechanisms within L2 and DRAM, is essential for optimizing applications, designing new architectures, and building accurate simulators. However, the latest GPU hardware from vendors like NVIDIA and AMD is still a black-box, making it challenging for researchers to understand the details of their design. 

In this paper, we introduce \projname{}, a methodology designed to unveil the NUMA architecture of the GPU memory hierarchy through microbenchmarking and data analysis. Specifically, we propose an approach to measuring the latency of L2 caches and DRAM without relying on the intrinsic instructions of the architecture and apply a Gaussian mixture model to filter out outliers and accurately determine latency distributions. 
We apply \projname{} on NVIDIA's A100 and H100 GPUs, revealing NUMA node architecture, SM-NUMA relationships, and NUMA-aware memory allocation strategies used to maintain cache coherence. 
To the best of our knowledge, this is the first paper to detail the NUMA architecture within the GPU memory subsystem.
\end{abstract}
\maketitle 
\let\thefootnote\relax\footnotetext{New Paper, Not an Extension of a Conference Paper.}

\section{Introduction}
GPUs have become essential across a broad range of fields, including gaming~\cite{park2020blackmirrorgpugaming,prakash2016improvinggpugaming}, graphics rendering~\cite{ren2021chopingpugraphicsrendering}, ray tracing~\cite{gu2024gvulkangpuraytracing}, and artificial intelligence~\cite{chen2018tvm,pytorch,abadi2016tensorflow}. Rapid advancements in GPU core development have enabled these processors to achieve remarkable performance levels, reaching up to 33.5 peak FP-64 TFLOPS~\cite{nvidia-h100}. 
However, while computational power has surged, GPU memory has not kept pace, creating a gap that can limit GPU workload efficiency.

Understanding GPU memory is essential for several reasons. First, it allows programmers to develop highly efficient code that maximizes the utilization of GPUs.
Second, it aids researchers in understanding the latest GPU designs, enabling the design of more sophisticated and optimized architectures.
Third, a deep understanding of GPU design is essential for developing accurate simulators.

However, a significant challenge lies in the closed-source nature of advanced GPUs from major vendors, including NVIDIA and AMD. While numerous prior works~\cite{dissectinghopperarchitecture,jia2019dissectingturing,jia2018dissectingvoltagpu,sun2023dissectingtensorcores} have proposed methods to reveal memory hierarchy details, including cache line size, hit latency and bandwidth, there is a lack of approaches aiming to reveal L2 and DRAM Non-Uniform Memory Access (NUMA) mechanisms on GPUs. In contrast, recent research works~\cite{li2023transgpunuma,wang2024gritgpunuma,milic2017beyondgpunuma,lee2023snakebytegpunuma,li2023orchestratedgpunuma,young2018combininggpunuma,agarwal2015unlockinggpunuma,DBLP:conf/isca/XieFCS19gpunuma,ren2020hmggpunuma} have focused on improving GPU NUMA architecture to optimize memory access patterns, minimize contention, and enhance scalability. To illustrate the methodologies employed by industry and to establish a solid baseline for researchers, approaches that provide insight into the NUMA architecture of GPUs are required.

In this paper, we introduce \projname{}, a methodology designed to reveal the NUMA architecture in GPU memory subsystems.
\projname{} measures L2 and DRAM latencies independently from vendor-provided intrinsic instructions, which we show can provide misleading results in specific instances. 
Then we develop a methodology to automatically filter out outliers, often triggered by factors such as DRAM refresh, and subsequently derive accurate and reliable memory latency values using a Gaussian mixture model~\cite{song2017multimodalgaussain,damianou2012manifoldmultimodalgaussian}.
\projname{} analyzes the distribution of memory accesses across different levels of the memory hierarchy and their NUMA nodes. Additionaly, by examining the NUMA architecture, \projname{} illustrates the correspondence between Graphics Processing Clusters (GPCs), GPU streaming multiprocessors (SMs), and L2 and DRAM NUMA nodes.  Furthermore, \projname{} unveils mechanisms of read and write instructions on NUMA nodes, providing deeper insights into the memory accesses. To demonstrate the broad applicability of \projname{} across various GPUs, we apply \projname{} to the latest NVIDIA GPU architectures, including Ampere (A100) and Hopper (H100).

First, our analysis reveals several key findings for the A100 and H100 GPUs based on the latency distribution. 
Using a Gaussian mixture model, we derive the latencies for reading data on local and remote DRAM NUMA nodes, which are \dramamperegpulow{} and \dramamperegpuhigh{} on the A100, and \dramhoppergpulow{} and \dramhoppergpuhigh{} on the H100, respectively.
Additionally, reading data on
local and remote L2 NUMA nodes results in latencies of \llcamperegpulow{} and \llcamperegpuhigh{} on the A100 and \llchoppergpulow{} and \llchoppergpuhigh{} on the H100, respectively. 

We observe that L2 NUMA nodes on both the H100 and A100 operate under a first-touch mechanism, where the latency for an SM accessing a data block is affected by previous SMs that accessed the same block. Furthermore, we also find that H100 incorporates sub-NUMA nodes~\cite{alappat2020understandingsubnuma} within all L2 NUMA nodes, introducing an access latency gap \subllchoppergpugap{} accessing sub-NUMA nodes. In contrast, the A100 lacks sub-NUMA nodes, which we attribute to the larger L2 cache size on the H100 (50 MB), compared to the A100 (40 MB). 

By analyzing latency gaps, we conclude that SMs access remote L2 and DRAM NUMA nodes through interconnections between L2 NUMA nodes as the observed NUMA latency gaps for accessing remote NUMA node of L2 and DRAM are similar.
Specifically, on the A100, we observe a NUMA latency gap of \dramamperegpugap{} for DRAM and \llcamperegpugap{} for L2 caches, while on the H100, the NUMA latency gaps on DRAM and L2 are \dramhoppergpugap{} and \llchoppergpugap{}, respectively.

Second, \projname{} reveals the corresponding relationship between SMs, GPCs, L2 NUMA nodes, and DRAM NUMA nodes. 
We find that 2 GPCs, each containing a group of SMs, connect to a sub-NUMA node, while 4 GPCs connect to an L2 NUMA node on both the A100 and H100. The distribution of SMs on the A100 is imbalanced across different L2 NUMA nodes, while the H100 addresses this imbalance for L2 NUMA nodes. However, H100 still exhibits imbalances when it comes to the sub-NUMA nodes.

Third, \projname{} reveals the read and write mechanisms of NUMA nodes on A100 and H100 GPUs.
The home node for a data block refers to the node that stores its values~\cite{dashti2013trafficnumahomenode}.
We define the \textit{DRAM home node} as the node that contains the physical page of the data block.
The \textit{L2 home node} is the node directly connected to the DRAM home node,  while the \textit{ home SM} refers to the SM node that is directly connected to the L2 home node. The terms \textit{DRAM remote node}, \textit{L2 remote node}, and \textit{remote SM} refer to the reverse of the home nodes.

Our findings indicate that for read operations, home SMs directly read the data block into the L2 home node, while remote SMs buffer data into both L2 home and remote NUMA nodes. For write operations, home SMs write data blocks solely to the L2 home node.
In contrast, remote SMs update both local and remote NUMA nodes if a cache hit occurs in the L2 home node. If the cache hit occurs only on the L2 remote node, remote SMs only update the L2 remote node.

In summary, we make the following contributions:
\begin{itemize}[leftmargin=*,nosep]
 \item We propose \projname{}, a framework to reveal the NUMA architecture of the GPU memory hierarchy through microbenchmarking and data analysis. We will
open-source \projname{} to support further research.
 \item We propose an independent latency measurement methodology. \projname{} accurately captures L2 and DRAM latencies without relying on vendor-specific intrinsic instructions, which can introduce misleading values at certain data points. We employ a mathematical approach, the Gaussian mixture model, to automatically detect and remove outliers caused by various factors across different architectures. By preprocessing the dataset, we can derive a more accurate understanding of memory access behavior.
 \item Our analysis reveals the corresponding relationship among SMs, L2 and DRAM NUMA nodes and the NUMA allocation mechanisms.
 We find that L2 NUMA nodes in both GPUs operate based on a first-touch mechanism, affecting latency depending on prior SM access. 
 Furthermore, we identify that the H100 includes sub-NUMA nodes within its L2 NUMA nodes, likely due to its larger cache size, while A100 lacks this sub-structure. 
 \item We demonstrate the read and write mechanisms of L2 NUMA on the A100 and H100.
 We conclude that home SMs perform direct reads and writes to home L2 nodes, while remote SMs apply specific buffering and updating strategies across both local and remote NUMA nodes. 
\end{itemize}

\section{Background}

\textbf{GPU memory hierarchy.}  The GPU memory hierarchy is structured to optimize data access and computational efficiency, beginning with registers at the core level to the slowest levels of GPU memory. GPU registers are used to store temporary data for warps, providing the quickest access to frequently used values during computation. Next, each SM contains a shared memory, which is a fast, user-managed memory, allowing high-speed data exchange between warps within the same thread block. Each SM also has its own L1 cache, a small, local SRAM, to provide fast access to frequently used data. On advanced GPUs like the A100 and H100, shared memory is dynamically allocated from the L1 cache according to kernel execution requirements.

The L2 cache is shared across all SMs and serves as a large, global cache, typically organized using NUMA architecture since L2 caches are significantly large on modern GPUs. The largest and slowest level of the memory hierarchy is global memory, typically organized using NUMA architecture and composed of high-bandwidth DRAM such as GDDR6, HBM2, or HBM3. 
However, vendors, including AMD and NVIDIA, do not provide detailed information about the L2 and DRAM NUMA architecture in their public documentation.


\textbf{Dissecting GPU Architecture.} 
Understanding GPU architecture is attracting more and more attention from researchers as dissecting the details of GPU design aids in fields, including improving GPU application performance and developing novel GPU architectures.
Luo et al~\cite{dissectinghopperarchitecture} introduce a benchmarking suite designed to unveil the tensor core performance and programming features of the Hopper architecture. Similar studies have been conducted on other GPU architectures, such as  Turing~\cite{jia2019dissectingturing}, Volta~\cite{jia2018dissectingvoltagpu}, and Ampere~\cite{sun2023dissectingtensorcores} architectures. However, these works do not demonstrate the details of the NUMA architecture and cache coherence protocol of GPUs, which is crucial for a comprehensive analysis GPU performances.

In addition to these works, GPU security researchers have also contributed on dissecting GPU architectures. For example, \textit{Spy in the GPU-box}~\cite{dutta2023spyinthegpubox} reveals that memory locations are cached only on the GPU where the memory is allocated, specifically in the Pascal and Volta GPU architectures. INVALIDATE+COMPARE~\cite{zhang2024invalidategpusecurity} uses cache maintenance instructions to reverse engineer some properties of the cache hierarchy. However, these works primarily focus on attack methodologies, with the dissected aspects of the GPU memory hierarchy serving to support their approach.
In contrast, \projname{} presents a comprehensive methodology for revealing the full details of the GPU NUMA architecture.


\textbf{NUMA structure and relative GPU works.} Non-Uniform Memory Access (NUMA) is commonly used in many-core systems where memory is divided into multiple nodes, each associated with one or more cores. 
In NUMA systems, processors can access their local memory more quickly than memory located on other nodes, leading to varying access times according to the distance of memory to the processor. 

The NUMA architecture in GPU memory has gained attention from researchers seeking to improve GPU performance. 
Prior works have explored the methodology to maintain the cache coherence~\cite{ren2020hmggpucoherence,alsop2016lazygpuconsistency,alsop2018spandexgpucoherence,dalmia2024cpelidegpucoherence,koukos2016buildinggpucoherence,alshboul2018lazypersistency} and improve the NUMA performance~\cite{ren2020hmggpunuma,li2023transgpunuma,wang2024gritgpunuma,xie2019oogpunumaperf,young2018combininggpunuma,milic2017beyondgpunuma,roy2018numagpunumaperf,arunkumar2017mcmgpunuma,lee2023snakebytegpunuma,zhang2015numagpunumaperf,li2023orchestratedgpunuma,agarwal2015unlockinggpunuma} on GPUs.
Zhao et al.~\cite{zhao2023nuba} propose the Non-Uniform Bandwidth Architecture to improve the bandwidth of the last-level cache.
Milic et al.~\cite{milic2017beyondgpunuma} propose NUMA-aware multi-socket GPUs to minimize the impact of NUMA effects. 
Runtime methodologies~\cite{agarwal2015unlockinggpunuma,wang2024gritgpunuma} reduce the NUMA overhead via optimizing the page placements.
A deeper understanding of the NUMA architecture in commercial GPUs can help researchers contribute to optimizing GPU NUMA systems.

\section{Motivation and Challenges}


Due to the high demands of GPU workloads, such as the training of large language model~\cite{sheng2023flexgengpullm,llamallm,achiam2023gpt}, GPU memory and caches are becoming larger and \tacorebuttal{larger}. Therefore, the introduction of NUMA architecture becomes inevitable due to factors, such as the physical limitations of a single node. However, NUMA introduces imbalanced access latencies between different NUMA nodes, and accessing remote NUMA nodes has a higher latency than that of the local NUMA node. 
Therefore, numerous works have been proposed to mitigate these NUMA overheads~\cite{ren2020hmggpunuma,li2023transgpunuma,wang2024gritgpunuma,milic2017beyondgpunuma,lee2023snakebytegpunuma,li2023orchestratedgpunuma,young2018combininggpunuma,agarwal2015unlockinggpunuma,dashti2013trafficnumahomenode,DBLP:conf/isca/XieFCS19gpunuma}. However, due to the black-box nature of commercial GPUs, the gap between prior NUMA research and the commercial implementation of NUMA in GPUs is implicit, which limits the progress of the community in this area.

A deep understanding of GPU NUMA architecture is essential across various fields. For example, researchers can leverage knowledge of GPU NUMA architecture to improve the performance of GPU workloads~\cite{xie2019oogpunumaperf,roy2018numagpunumaperf,zhang2015numagpunumaperf}.
Furthermore, researchers focusing on GPU NUMA architecture~\cite{ren2020hmggpunuma,li2023transgpunuma,wang2024gritgpunuma,milic2017beyondgpunuma,lee2023snakebytegpunuma,li2023orchestratedgpunuma,young2018combininggpunuma,agarwal2015unlockinggpunuma,DBLP:conf/isca/XieFCS19gpunuma} can advance their work based on insights into commercial GPUs. Additionally, GPU simulator~\cite{sun2019mgpusim,khairy2020accel} can keep pace with the latest GPUs and provide an accurate foundation for researchers to build on. GPU security~\cite{dutta2023spyinthegpubox,zhang2024invalidategpusecurity} researchers also require hardware knowledge to design their attack and defense mechanisms.


However, dissecting GPU NUMA architecture poses significant challenges. Modern GPUs feature complex memory hierarchies and virtual memory subsystems, which complicate isolating and testing individual components. For example, to accurately measure L2 cache latency, it is necessary to account for and minimize the influence of memory address translation overhead. Moreover, the data required to be fetched from DRAM into L2 should bypass L1 caches to isolate the L2 cache's specific latency.

\begin{table}
    \centering
    \caption{The corresponding relationship among CUDA code, PTX and SASS instructions}
    \begin{tabular}{ccc}
    \toprule
        CUDA code & PTX & SASS (A100, H100) \\
    \midrule
        \_\_ldcg(ptr) & ld.global.cg.u8 & LDG.E.U8.STRONG.GPU\\
        *ptr&ld.global.u8&LDG.E.U8\\
    \bottomrule
    \end{tabular}
    \label{tab:ldcgptxsass}
\end{table}

\begin{figure}
    \centering
    \begin{subfigure}{0.49\linewidth}
    \includegraphics[width=\linewidth]{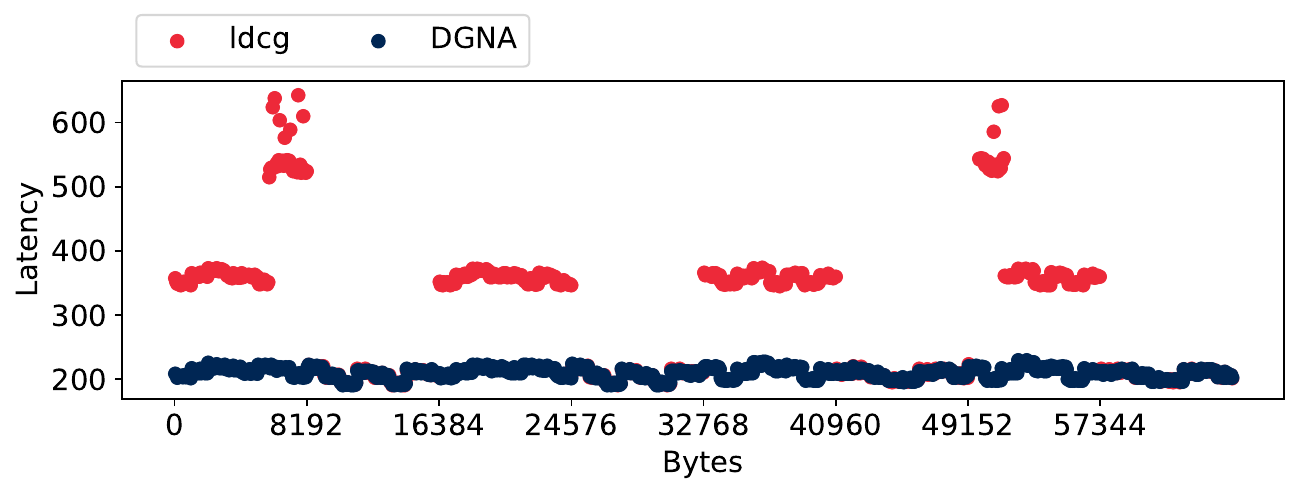}
    \caption{A100}
    \label{fig:ldcg:a100}
    \end{subfigure}
    \begin{subfigure}{0.49\linewidth}
    \includegraphics[width=\linewidth]{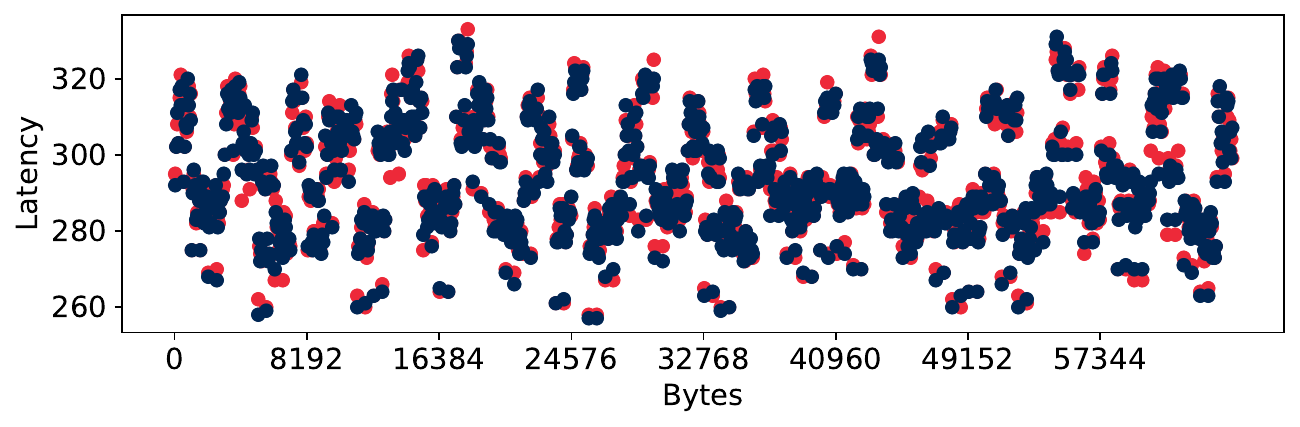}
    \caption{H100}
    \label{fig:ldcg:h100}
    \end{subfigure}
    \caption{The L2 latencies measured by \textit{\_\_ldcg}~\cite{dutta2023spyinthegpubox,jin2024uncoveringgpunoc} and \projname{} on the A100 show clear gaps, suggesting that \textit{\_\_ldcg} introduces additional operations for certain L2 memory accesses. For the H100, the similar results indicate that \textit{\_\_ldcg} may directly fetch data from the L2 cache. These findings suggest that vendor-provided intrinsics 
    contain undisclosed implementation details that can impact final results.}
    \label{fig:ldcg}
\end{figure}

Prior works, such as \textit{spy in the GPU-box}~\cite{dutta2023spyinthegpubox} and \textit{uncovering real GPU NoC characteristics}~\cite{jin2024uncoveringgpunoc}, rely on specialized load primitives, such as 
$\_\_ldcg()$, to constrain data accesses to L2 cache and thereby measure L2 latency. 
However, these vendor-specific primitives may not be supported on other GPU architectures.
Moreover, the PTX and SASS instructions generated by $\_\_ldcg()$ differ from those of ordinary memory accesses, as shown in \autoref{tab:ldcgptxsass}. Consequently, using the memory instructions generated by $\_\_ldcg()$ to analyze the behavior of ordinary memory accesses has inherent limitations,
raising questions about the exact feature being measured. 

The results, as shown in \autoref{fig:ldcg}, indicate that the latencies measured by \projname{} and \textit{\_\_ldcg}~\cite{dutta2023spyinthegpubox} differ on the A100 but are similar on the H100. The difference in latency measurements between \textit{\_\_ldcg} and ordinary instructions lies in the instructions shown in \autoref{tab:ldcgptxsass}, with all other is the same.
The L2 latencies of \projname{} are consistently around 200 cycles as these data blocks are fetched into the local L2 NUMA node by another SM on A100. However, with \textit{\_\_ldcg}, 
some latencies are concentrated around 350 cycles and some others are around 500 cycles, indicating that these data blocks are either cached in a different L2 NUMA node or still reside in DRAM.
We conclude that utilizing ordinary memory instructions offers more reliability.
%

Moreover, the latency variance in GPU DRAM and L2 cache subsystems can further complicate analysis.
For example, DRAM cells require periodic refreshing~\cite{nair2013casedramrefreshing,mukundan2013understandingrefreshing} to maintain data integrity which can interfere with latency measurements and lead to inaccurate conclusions. The DRAM page policy of opening and closing pages also affects the results. Additionally, factors such as TLB misses, temperature fluctuations, and power states of GPUs also introduce noise, skewing latency measurements. 
To accurately analyze GPU NUMA architectures, a general methodology is required to filter out outliers and ensure reliable results.


\section{Methodology}
\begin{listing}

\begin{minted}[escapeinside=||]{c}
#define LDCG_TEST_READ 1
#define TEST_READ 2
#define TEST_WRITE 3
template <int MODE>
__global__ void cuTestLat(int smid, int *smSet,int *dO, char * dMem) {
    unsigned int cur_smid;
    __shared__ unsigned char shared[1];
    asm("mov.u32 %0, %smid;" : "=r"(cur_smid));
    //ensuring only executing on the SM ("smid")
    if (smid != cur_smid) return;
    //ensuring only one warp from a single thread block executing
    int old = atomicAdd(&smSet[smid], 1)
    if ( old != 0 ) return;
    if (threadIdx.x == 0) {
        uint64_t start = clock();
        if (MODE == LDCG_TEST_LOAD) {
            shared[0]+=__ldcg(dMem[0]);
        } else if (MODE == TEST_LOAD) {
            shared[0] += dMem[0];
        } else if (MODE == TEST_WRITE) {
            dMem[0] = 1;
            __threadfence();
        }
        dO[0] = clock() - start;
}}
int *dO, *dMem, *smSet;
template<int MODE>
int TestLatency(int smid, int idx) {
    int lat;
    int * addr = &dMem[idx]
    cudaMemset(smSet, 0, NUM_of_SM*sizeof(int));
    cuTestLat<MODE><<<1024,32>>>(smid,smSet,dO,addr);
    cudaMemcpy(&lat, dO, 4, cudaMemcpyDeviceToHost);
    return lat;
}
template<int MODE>
void TestLatencies( vector<int> &l2latencies,  vector<int> &dramlatencies, vector<int> sms) {
    int step = max(DRAMtoL2TransSize,L1CacheLineSize);
    for (auto sm_1 : sms) {
        for (auto sm_2 : sms) {
            malloc(d0, dMem, smSet)
            for (int i = 0 ; i < Size ; i+=step ) {
                //measure DRAM latency and warmup L2.
                int lat1 = TestLatency<MODE>(sm_1,i);
                //measure L2 latency.
                int lat2 = TestLatency<MODE>(sm_2,i);
                dramlatencies.push_back(lat1);
                l2latencies.push_back(lat2);
            }
            cudaDeviceReset();
}}}
\end{minted}
\caption{The code that measures the latency of accessing one byte in L2 cache and DRAM. \tacorebuttal{We use the \texttt{malloc} function to allocate GPU memory after resetting the device, ensuring that each measurement does not affect the others.}
}
\label{listing:testl2:code}
\end{listing}

In this section, we present the methodology used to uncover the complexity of GPU NUMA architecture. First, we describe how to measure L2 cache and DRAM latencies without relying on specific intrinsic instructions. Next, we introduce the approach to dissect the NUMA architecture topology. Finally, we explain the techniques used to reveal the read and write mechanisms on GPU NUMA architecture.

\tacorebuttal{DGNA assumes that GPUs do not employ hardware prefetching. We argue that future commercial GPUs are unlikely to include prefetchers for several reasons. First, a GPU can be viewed as an execution-driven prefetcher as massive thread-level parallelism allows it to issue memory requests for future accesses early enough to hide DRAM latency, thereby reducing the need for complex prefetching hardware, which is commonly used in CPUs. 
Second, adding prefetching hardware would increase complexity and reduce the number of cores that can fit on a given die size. Finally, modern GPUs are often bandwidth-limited, and prefetching can introduce unnecessary memory traffic, potentially degrading performance rather than improving it~\cite{bera2022hermes,pakalapati2020bouquet}. Rather than using prefetchers, some GPUs employ the Tensor Memory Accelerator (TMA)~\cite{choquette2022nvidia}, a specialized unit that asynchronously issues strided and tiled memory accesses, to improve bandwidth utilization.}


\subsection{Measuring Latency}
\begin{figure}[t]
    \centering
    \includegraphics[width=.39\linewidth]{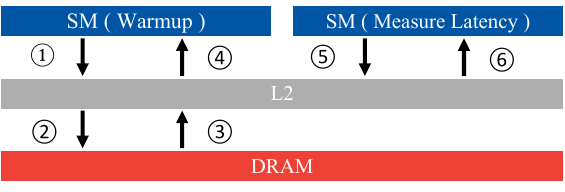}
    \caption{\projname{} leverages the fact that L2 caches are shared among all SMs, while each SM has its own private L1 cache, to measure L2 latency. \projname{} uses one SM to warm up the L2 cache and another SM to measure the L2 latency, as the L1 caches are not shared between these two SMs.}
    \label{fig:method:latency}
\end{figure}
Measuring latencies for GPU NUMA architecture poses a significant challenge, as vendors do not provide specific instructions that allow programmers to control which NUMA nodes and memory levels are accessed by memory instructions.
Although vendors, such as NVIDIA and AMD, provide intrinsics intended to access data at particular memory levels, the exact hardware implementations of these instructions remain undisclosed, leaving researchers uncertain about the mechanisms underlying them.


Instead of relying on vendor-specific intrinsic instructions, \projname{} introduces a new methodology to measure latencies for each level of the memory hierarchy independently of these specific instructions. \autoref{fig:method:latency} illustrates how \projname{} measures the latencies of L2 and DRAM when accessing the L2 cache and DRAM. First, \projname{} employs one SM (\textit{SM (Warmup)}) to fetch data from DRAM into both the L2 cache and its own L1 cache. Since the L2 cache is shared across all SMs, the data resides in L2. \projname{} then uses a different SM (\textit{SM (Measuring Latency)}) to issue memory requests and measure the latency of accessing the L2 cache, as each SM has its own private L1 cache.

List~\ref{listing:testl2:code} presents the code used to analyze the L2 and DRAM latencies. Specifically, \projname{} uses one SM to load a data block into the L2 cache and a different SM to subsequently fetch this data block from L2, as each SM maintains its own L1 cache, isolating the access paths and ensuring accurate measurement of latency between cache levels.

The function \textit{cuTestLat}, running on GPUs, measures the memory access latency for the memory address \textit{d\_mem} on the particular SM (\textit{smid}). 
Given that GPU schedulers cannot be directly controlled by programmers, 
\tacorebuttal{\projname{} launches a large number of thread blocks (2048 in our experiments), each containing a single warp. It then filters out thread blocks that are not scheduled on the target SM (Lines 6–10). Because multiple thread blocks may still be scheduled on the same SM, a counter (Lines 12–13) is used to restrict measurements to a single warp from a single thread block. }
Only thread 0 within the warp reads/writes data (Lines 16-23), which prevents memory requests from other threads from interfering with latency measurements. The fetched value is added to shared memory (\textit{shared}) or \textit{\_\_threadfence} is used to ensure 
the memory access fully completes.

The function \textit{TestLatency}, running on the CPU, resets the warp counter for each SM (Line 34) before launching \textit{cuTestLat} to measure memory access latency. The \textit{MODE} parameter specifies whether the measurement mode is for read, write, or read operations through vendor-specified intrinsics.

The function \textit{TestLatencies} measures memory latency values across all pairwise combinations of SMs for a sequence of memory addresses as each factor influences latency results. 
The first call to \textit{TestLatency} in \textit{TestLatencies} (Line 45) measures the latency of fetching data from DRAM and warmup the L2 cache, while the second call (Line 47) specifically measures the L2 cache latencies.
The function \textit{cudaDeviceReset} is called when either \textit{sm\_1} or \textit{sm\_2} changes to prevent prior measurements from affecting current results. 
We do not reset the device for each new address to accelerate the measurement process as,
based on our observations, 
the A100 and H100 GPUs lack prefetch mechanisms. 

After collecting the dataset using \projname{}, we preprocess the data to
filter out outliers caused by factors, such as TLB misses, DRAM refreshing. Assume that latencies are only influenced by NUMA architecture, the latency distribution can be modeled as a Gaussian mixture, given
by 
\vspace{-5pt}
\begin{equation}
p(\mathbf{lat}) = \sum_{i=1}^{K} \pi_i \mathcal{N}(\mathbf{lat} | \mu_i, \Sigma_i)
\end{equation}
where $lat$ represents the latencies, $p(lat)$ is the probability density function, $K$ is the number of Gaussian components in the mixture, $\pi_i$ denotes the proportion of latencies associated with each NUMA node's data block access by SMs, $\mu_i$ is the mean latencies for each component, and $\Sigma_i$ is the variance for each Gaussian component.

Since we assume that NUMA architecture is the primary factor influencing latencies, the means $\mu_i$ represent the latencies of different SMs accessing various NUMA nodes, and these may differ. 
However, we expect the variances $\Sigma_i$ to be similar across different components, as NUMA architecture is assumed to be the main latency factor. 
During the analysis, we observe that \(\Sigma_0\) (the variance for SMs accessing their local NUMA nodes) remains stable, while the variances \(\Sigma_i\) for \(i > 0\) (representing remote accesses) are more unstable. This is because other factors tend to increase latencies for memory accesses. Therefore, we use \(\Sigma_0\) as an approximation for the \(\Sigma\) in other Gaussian components.

Since the probability of a Gaussian distribution for values beyond \(3 \times \Sigma\) is small, we remove the outliers by filtering latencies that are larger than $\mu_K+3\times \Sigma$ since outliers tend to be very large.
After this filtering, we obtain the final latency dataset. 
This Gaussian mixture model (GMM) separates latencies from SMs accessing different NUMA nodes, providing a framework to filter out outliers without needing knowledge of varying factors, such as TBL misses, DRAM refreshing, that can differ across GPUs. 

\tacorebuttal{When scaling beyond two NUMA nodes, the latency distribution may become multimodal rather than bimodal, potentially with higher variance. A potential solution can overcome the higher variance through the following steps. First, remove outliers for each address to reduce measurement noise. Second, combine the filtered measurements into a latency vector, where each dimension corresponds to an address-specific latency. In the high-dimensional latency space, memory regions mapped to different NUMA nodes naturally form distinct clusters due to their structural latency differences. Finally, apply a random projection to map the latency vectors into a scalar space, allowing the GMM to remove the outliers further. }

\subsection{NUMA Architecture Topology}
\label{sec:method:memorytopo}
\begin{figure}[t]
    \centering
    \includegraphics[width=0.5\linewidth]{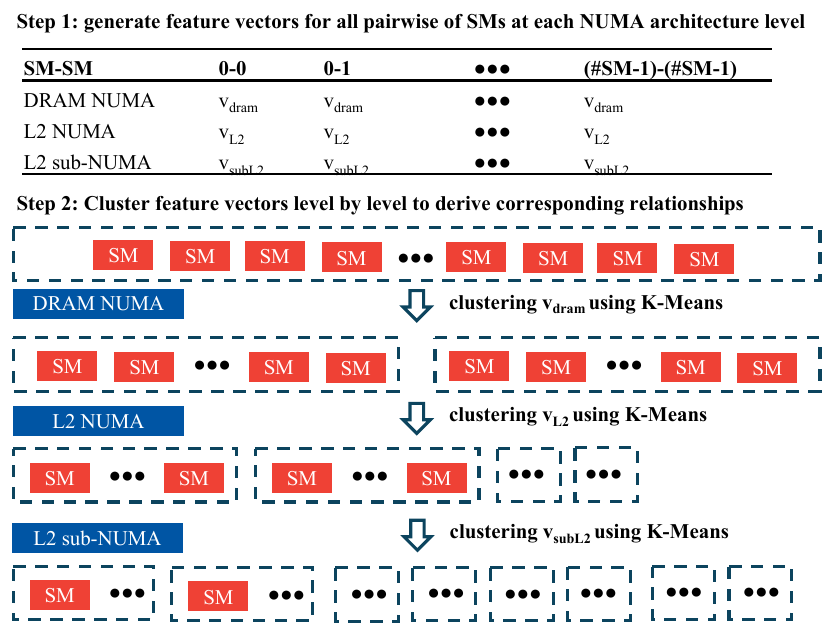}
    \caption{The workflow for analyzing the NUMA architecture topology. The first step is to generate feature vectors for any pair of SMs accessing data blocks across a sequence of memory addresses at different NUMA levels: DRAM NUMA, L2 NUMA and L2 sub-NUMA. 
    The feature vectors consist of latency values for accessing different memory addresses.
    The second step is to cluster the feature vectors obtained from DRAM NUMA to L2 sub-NUMA. It maps out the connections within the NUMA architecture, establishing which SMs are associated with specific DRAM, L2 and sub-L2 NUMA nodes.}
    \label{fig:method:topology}
\end{figure}

\projname{} first investigates the NUMA architecture hierarchy within the GPU by analyzing the statistical distribution of memory latency values.  We derive the latencies of accessing home and remote NUMA nodes and the possibility of sub-NUMA nodes~\cite{alappat2020understandingsubnuma} within the DRAM and L2 NUMA node, based on the Gaussian mixture model. \tacorebuttal{We observe that the mapping between allocated virtual pages and NUMA nodes remains stable after resetting the device (Line 50 in Listing~\ref{listing:testl2:code}). We leverage this stable mapping to remove the effects of virtual memory, thereby unveiling the details of the NUMA architecture. The details of this mapping are illustrated in Section~\ref{sec:eval:numaarchtopology}.}

Next, we leverage the latency differences when accessing various DRAM, L2 NUMA and L2 sub-NUMA nodes to analyze the GPU memory topology. \autoref{fig:method:topology} illustrates the workflow of \projname{} for analyzing NUMA architecture topology. We first generate feature vectors, which consist of latency values for accessing a sequence of memory addresses with all SM pairs.

Then, \projname{} first groups SMs based on the access latencies to DRAM NUMA nodes. This initial grouping is performed since DRAM latencies by each SM can be directly measured, while L2 NUMA nodes that buffer data depend on the SMs that initially access the data, complicating the relationship analysis between SMs and L2. 

Next, \projname{} groups SMs with their corresponding L2 NUMA nodes, further dividing these SMs into smaller groups. This is feasible as SM pairs on the same local NUMA node exhibit different latency characteristics compared to SM pairs on different L2 NUMA nodes.
Additionally, for GPUs with sub-NUMA architecture, \projname{} further splits SMs into groups corresponding to L2 sub-NUMA nodes~\cite{alappat2020understandingsubnuma}. \tacorebuttal{As observed, the L2 sub-NUMA on the H100 employs a memory interleaving scheme, in which data is mapped to different sub-NUMA nodes according to their addresses. By comparing L2 access latencies from different SMs, as measured using the methodology in Listing~\autoref{listing:testl2:code}, we can reconstruct the topology of SMs relative to the L2 sub-NUMA nodes.}


\projname{} uses the K-Means algorithm~\cite{ahmed2020kkmeans} to automatically analyze and cluster these feature vectors. 
By incorporating feature vectors that consist of latency measurements across a sequence of memory addresses, \projname{} helps mitigate the impact of this variance, enabling more accurate clustering.

\tacorebuttal{Although the current hardware exposes only two NUMA nodes, our methodology is generic and not restricted to this configuration. We agree that future GPUs may expose more than two NUMA nodes, driven by the increasing demand for memory capacity and cache buffering as GPU workloads continue to scale. For environments with multiple NUMA nodes, DGNA can be naturally extended by using latency vectors instead of scalar latency measurements. Specifically, we can first measure a vector of access latencies from different addresses, which are mapped to different NUMA nodes. Next, we could project the latency vectors into a scalar space using a random projection matrix. Utilizing this methodology, DGNA could be extended to scenarios with more than two NUMA nodes without fundamental changes.}


\subsection{Read and Write Mechanism on L2 NUMA}

The read mechanism on GPUs can be divided into four cases, categorized by the types of SMs used to warm up and test latencies of L2 caches. For a given data block, we define two types of SMs, including
(1) \textit{Home-SM}, which is directly connected to the L2 NUMA node, which in turn is directly connected to the DRAM containing the physical page of the data block. 
(2). \textit{Remote-SM.}, which are not Home-SM for the specific data block.
Therefore, we categorize these four types by leveraging Home-SM and Remote-SM for warm-up and measuring the L2 NUMA nodes.



\begin{listing}
\begin{minted}[escapeinside=||]{c}
#define WARMUP_HOME 1
#define WARMUP_REMOTE 2
#define WARMUP_BOTH 3
#define READ_HOME 1
#define READ_REMOTE 2
#define WRITE_HOME 1
#define WRITE_REMOTE 2
template<int MODE>
void WarmUP() {
    if (MODE == WARMUP_HOME) {
        TestLatency<TEST_READ>(sm_home,0);
    } else if (MODE == WARMUP_REMOTE) {
        TestLatency<TEST_READ>(sm_remote,0);
    } else if (MODE == WARMUP_BOTH) {
        TestLatency<TEST_READ>(sm_home,0);
        TestLatency<TEST_READ>(sm_remote,0);
}}
template<int MODE>
void TestWrite() {
    if (MODE == WRITE_HOME) {
        TestLatency<TEST_WRITE>(sm_home,0);
    } else if (MODE == WRITE_REMOTE){
        TestLatency<TEST_WRITE>(sm_remote,0);
}}
template<int MODE>
void ReadAfterWrite() {
    if (MODE == READ_HOME) {
        TestLatency<TEST_READ>(sm_home,0);
    } else if (MODE == READ_REMOTE) {
        TestLatency<TEST_READ>(sm_remote,0);
}}
template<int WARMMODE, int WRITEMODE, int READMODE>
void Test(int &lat_write, int &lat_raw) {
    Warmup<WARMMODE>();
    lat_write = TestWrite<WRITEMODE>();
    lat_raw = ReadAfterWrite<READMODE>();
}
\end{minted}
\caption{The code to dissect the write mechanisms within GPU's NUMA architecture. 
}
\label{listing:testwrite:code}
\end{listing}
Then we analyze the write mechanism on GPUs. As shown in Listing~\autoref{listing:testwrite:code}, \projname{} tests 12 different cases to measure the write mechanisms. The process involves three steps. (1). L2 cache warmup. \projname{} begins by warming up the L2 caches in 3 possible cases: home-SM, remote-SM, or both. (2). Write latency measurement. Once the L2 cache is warmed up, 
\projname{} measures the latencies of writing data to the same memory location. The write operation is tested using two cases: home-SM write or remote-SM write. A longer write latency generally indicates that more NUMA nodes are being updated during the write operation compared to a shorter write latency.  3). Read latency measurement after write (RAW). After the write operation, \projname{} tests the latencies of reading the data to analyze which NUMA nodes are updated during the write. 
Short read latencies indicate that the current NUMA node being accessed is already up to date.
All these steps help dissect the impact of write operations on the NUMA architecture, revealing how and where data is updated within the GPU memory subsystem.

\section{Evaluation}
In this section, we describe the findings on the A100 and H100. using \projname{}.

\subsection{Experimental Setup}
We evaluate \projname{} on two modern GPU architectures, A100~\cite{nvidia-a100} and H100~\cite{nvidia-h100}. H100 is the most advanced GPU architecture that can be purchased at the time. The configuration details of A100 and H100 are shown in ~\autoref{tab:arch:configs}. Both GPUs are equipped with 5 HBM stacks, with each stack having 2 memory controllers. Furthermore, both A100 and H100 have 2 L2 NUMA nodes. 
The transaction size from L2 to L1 is 32B on both the A100 and H100. 
NVIDIA provides an API to configure the default transaction size from DRAM to L2, and for this evaluation, we set it to the default value of 64B on both A100 and H100~\cite{thomas2020optimizingamperegpu}.
\begin{table}[h]
    \centering
    \caption{The architecture configurations for the GPUs evaluated, including A100~\cite{nvidia-a100} and H100 with SXM5~\cite{nvidia-h100}.}
    \label{tab:arch:configs}
    \begin{tabular}{c|cc}
        \toprule
        GPUs&A100 & H100 with SXM5 \\
        \midrule
        Architecture& Ampere & Hopper \\
        GPC & 7& 7 or 8\\
        SM $\times$ cores/SM & $108 \times 64$ & $132\times128$ \\
        L2  &40MiB&50MiB\\     
        L2 NUMA nodes &2&2 \\
        Memory &40GiB HBM2&80GiB HBM3\\
        HBM stack&5 &5\\
        Memory Controller &10 512-bit&10 512-bit\\
        \bottomrule
    \end{tabular}
\end{table}

\begin{figure}
    \centering
    \begin{subfigure}{0.49\linewidth}
    
    \includegraphics[width=\linewidth]{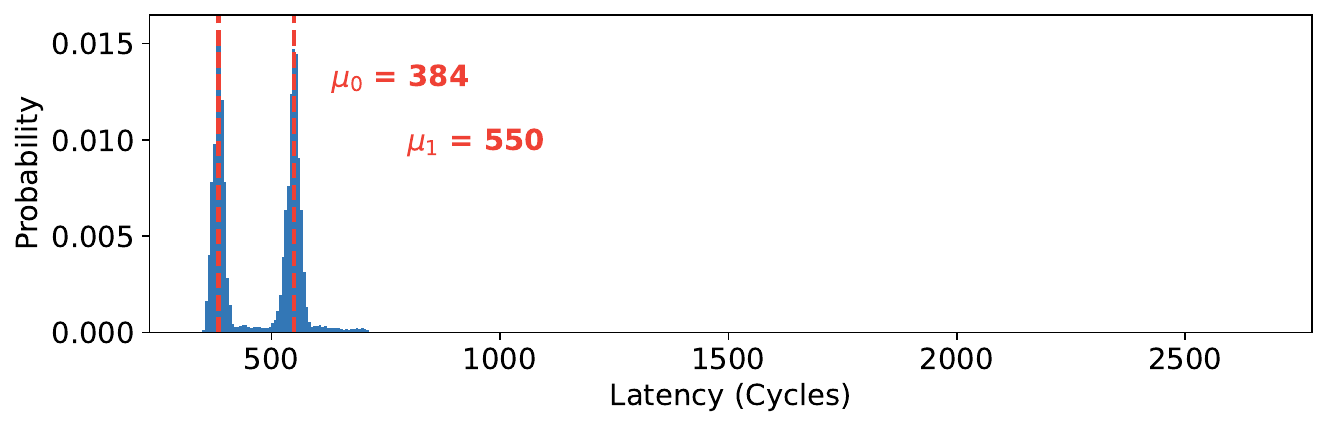}
    \caption{Original data on the A100}
    \label{fig:dramlatency:allsm:a100:raw}
    \end{subfigure}
    \begin{subfigure}{0.49\linewidth}
    \includegraphics[width=\linewidth]{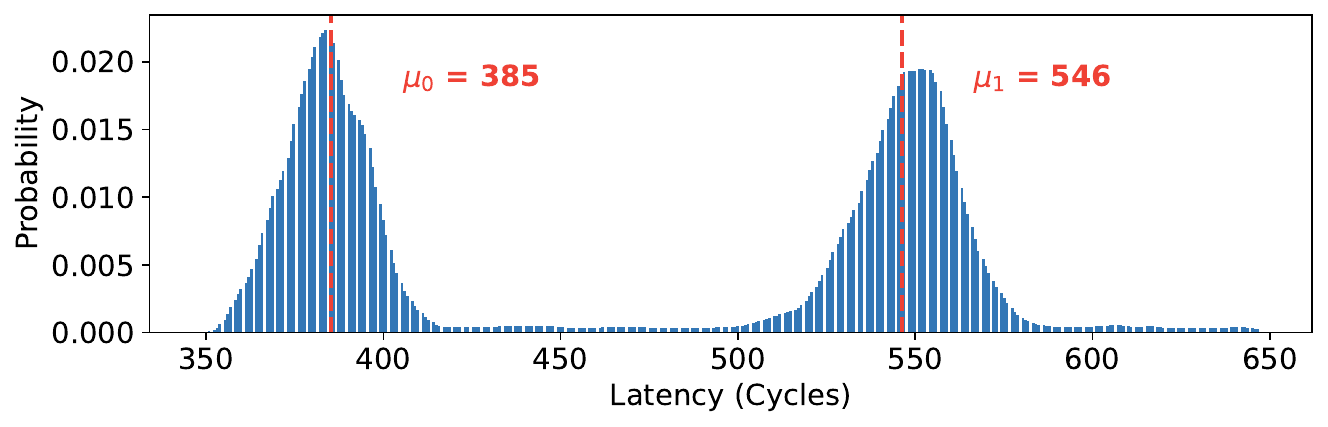}
    \caption{Data removed outliers on the A100 }
    \label{fig:dramlatency:allsm:a100}
    \end{subfigure}
    \begin{subfigure}{0.49\linewidth}
    \includegraphics[width=\linewidth]{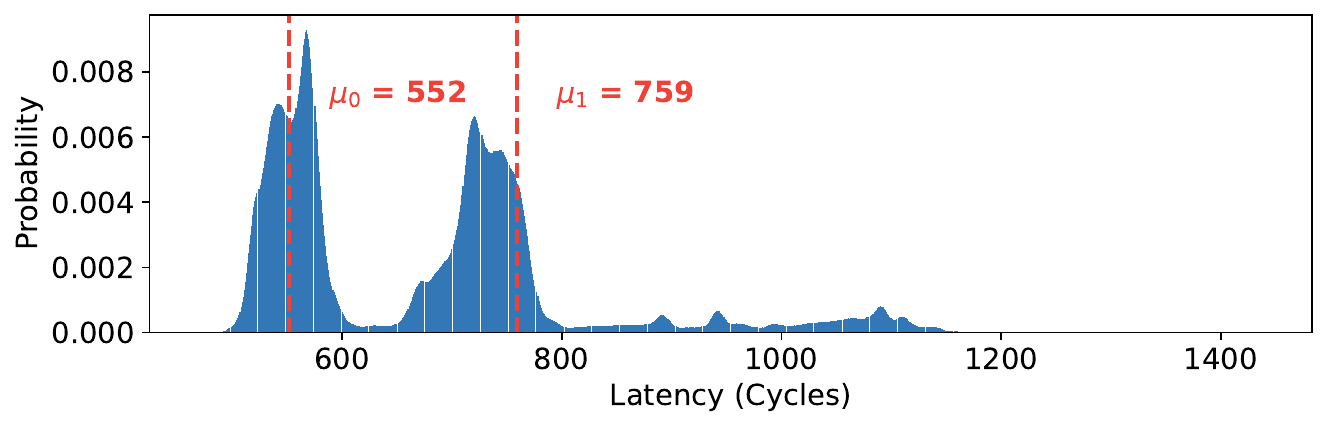}
    \caption{Original data on the H100}
    \label{fig:dramlatency:allsm:h100:raw}
    \end{subfigure}
      \begin{subfigure}{0.49\linewidth}
    \includegraphics[width=\linewidth]{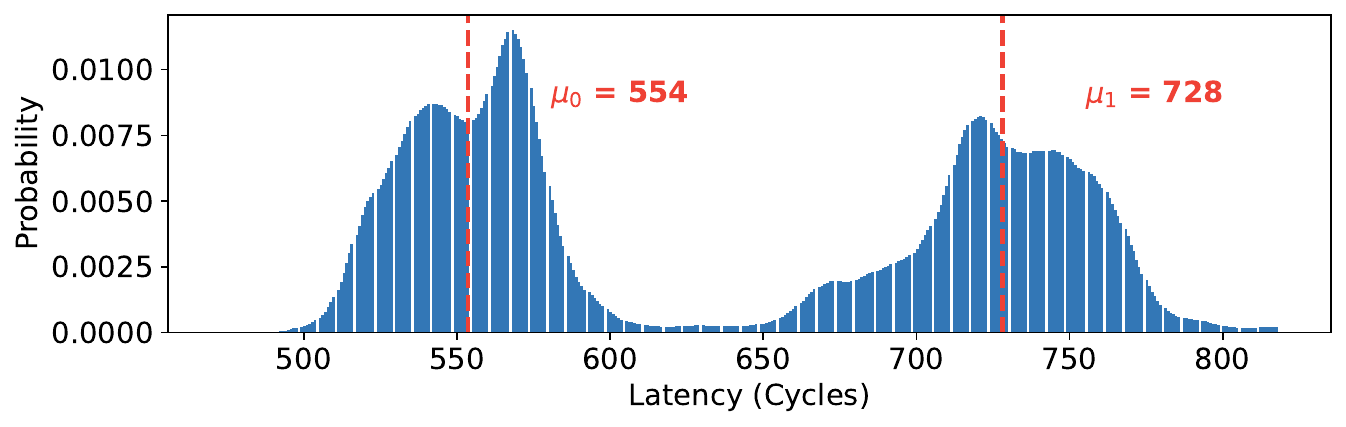}
    \caption{Data removed outliers on the H100 }
    \label{fig:dramlatency:allsm:h100}
    \end{subfigure}
    \caption{The distribution of access latencies for all SMs on GPUs accessing all DRAM NUMA nodes follows a mixed Gaussian distribution for both the A100 and H100 GPUs. These latencies are clusterred around two central values: \dramamperegpulow{} ($\mu_0$) and \dramamperegpuhigh{} ($\mu_1$) for the A100, and \dramhoppergpulow{} ($\mu_0$) and \dramhoppergpuhigh{} ($\mu_1$) for the H100 after removing outliers.}
    \label{fig:dramlatency:allsm}
\end{figure}

% ...

\subsection{Results of Measuring Latencies}
We measures latencies for accessing DRAM and L2 caches. We first demonstrate the distribution of raw data and data after removing outliers on DRAM latencies. Then we analyze the data distribution and give the the average latencies on DRAM, L2 NUMA nodes and L2 sub-NUMA nodes. 

\noindent\textbf{Filtering out outliers.} \autoref{fig:dramlatency:allsm:a100:raw} and \autoref{fig:dramlatency:allsm:h100:raw} illustrate the raw latency distributions for DRAM accesses on the A100 and H100, respectively. After filtering out outliers, the refined datasets are shown in \autoref{fig:dramlatency:allsm:a100} and \autoref{fig:dramlatency:allsm:h100}. The raw data on the H100, for example, exhibits some extreme values, leading to a distortion in the mean latencies. These outliers significantly shift the second mean to a higher value, \dramhoppergpuhighraw{}. Although the percentage of outliers is small, their impact on the values is significant. By filtering out these outliers, we obtain more accurate measurements, revealing the true means of accessing each DRAM NUMA node. 

\noindent\textbf{DRAM latencies.} 
The DRAM latencies on GPUs exhibit a mixed Gaussian distribution, with two distinct means observed for each architecture: \dramamperegpulow{} and \dramamperegpuhigh{} for the A100, and \dramhoppergpulow{} and \dramhoppergpuhigh{} for the H100.  The mixed Gaussian distribution with two means occurs because, despite the architectural differences, both the Ampere and Hopper GPUs employ a structure 
with 2 L2 NUMA nodes. Accessing DRAM through a local L2 NUMA node results in lower latencies, while accessing it through a remote L2 NUMA node incurs higher latencies. 

\begin{figure}[t]
    \centering
    \begin{subfigure}{0.49\linewidth}
    \includegraphics[width=\linewidth]{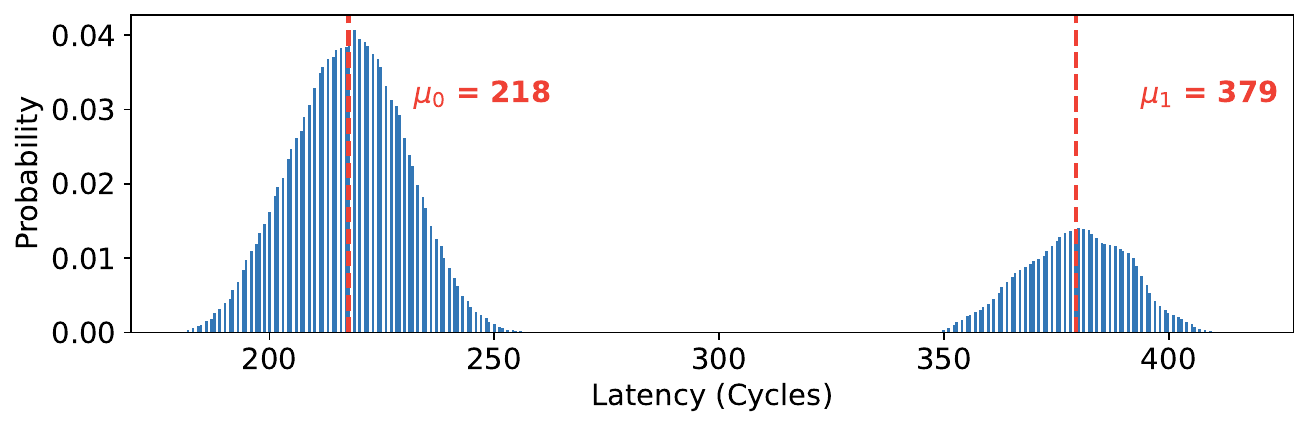}
    \caption{A100}
    \label{fig:l2latency:allsm:a100}
    \end{subfigure}
      \begin{subfigure}{0.49\linewidth}
    \includegraphics[width=\linewidth]{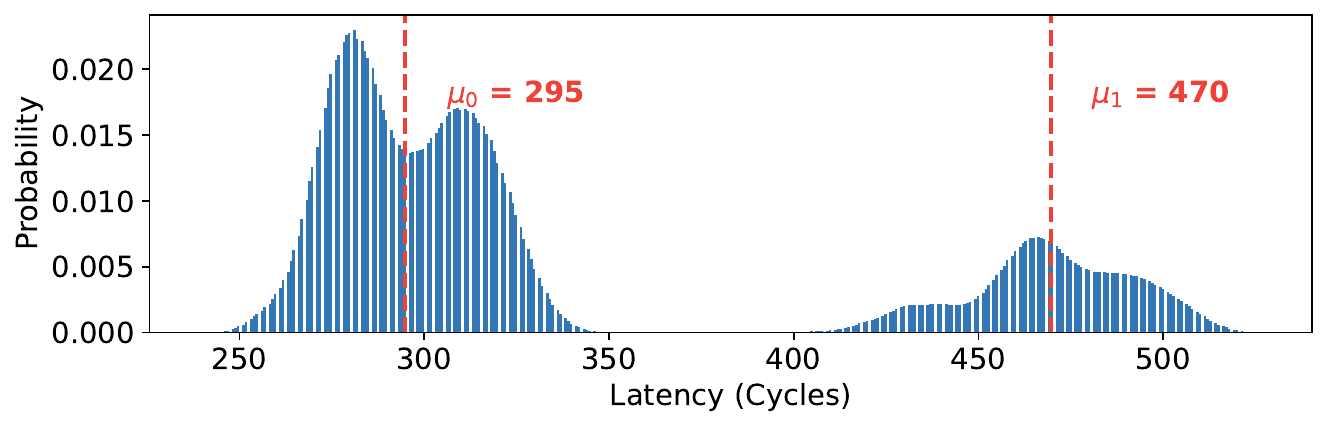}
    \caption{H100}
    \label{fig:l2latency:allsm:h100}
    \end{subfigure}
    \caption{The latency distribution for all SMs accessing all L2 NUMA nodes, after removing outliers, follows a mixed Gaussian distribution on both the A100 and H100 GPUs. The access cycles cluster around two central values: \llcamperegpulow{} ($\mu_0$) and \llcamperegpuhigh{} ($\mu_1$) for the A100, and \llchoppergpulow{} ($\mu_0$) and \llchoppergpuhigh{} ($\mu_1$) for the H100. The distribution of latencies on the H100 demonstrates that there are sub-NUMA nodes within the local NUMA node.}
    \label{fig:l2latency:allsm}
\end{figure}

\noindent\textbf{L2 latencies.} In \autoref{fig:l2latency:allsm:a100}, we observe that the latency distribution, after removing outliers, for the A100 follows a mixed Gaussian distribution, whose latencies are around \llcamperegpulow{} and \llcamperegpuhigh{}. For the H100, however, the latency distribution diverges from that of the A100, suggesting the presence of sub-NUMA nodes within the L2 cache. 
The average latencies for accessing local and remote NUMA nodes of the L2 cache on the H100 are \llchoppergpulow{} and \llchoppergpuhigh{}, respectively.

\begin{figure}[t]
    \centering
    \begin{subfigure}{0.49\linewidth}
    \includegraphics[width=\linewidth]{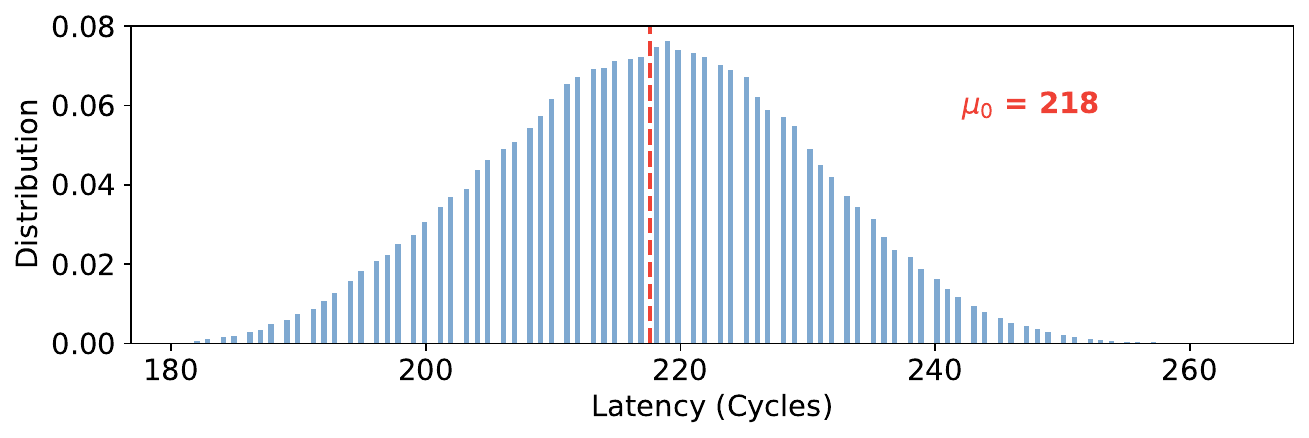}
      \caption{A100}
    \label{fig:l2latency:l2localnuma:a100}
    \end{subfigure}
    \begin{subfigure}{0.49\linewidth}
    \includegraphics[width=\linewidth]{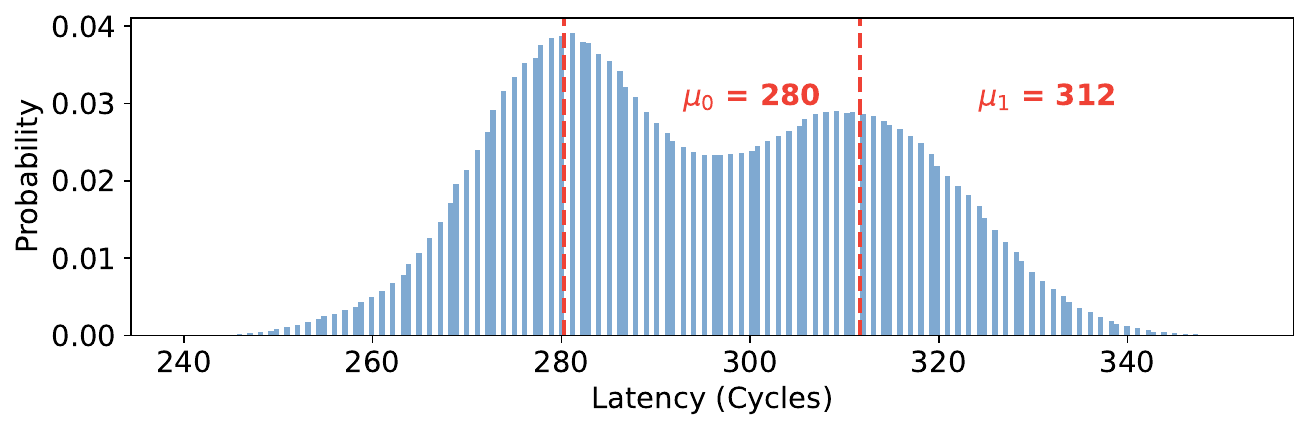}
     \caption{H100}
    \label{fig:l2latency:l2localnuma:h100}
    \end{subfigure}
    \caption{The latency distribution for all SMs accessing their local L2 NUMA nodes on the H100 follows a mixed Gaussian distribution, while on the A100 follows a Gaussian distribution. The mean latency values are \subllchoppergpulow{} ($\mu_0$) and \subllchoppergpuhigh{} ($\mu_1$), respectively, representing the two main latency groups within the local L2 NUMA access pattern for L2 sub-NUMA nodes on the H100. }
    \label{fig:l2latency:l2localnuma}
\end{figure}

\noindent\textbf{L2 sub-NUMA latencies}
On the H100, L2 latencies when accessing the local NUMA node show a mixed Gaussian distribution, while the A100 follows a single Gaussian distribution, 
as shown in \autoref{fig:l2latency:l2localnuma}. 
For the H100, the memory latency values exhibit two mean values, \subllchoppergpulow{} and \subllchoppergpuhigh{}. We attribute this to the larger L2 cache size of the H100 (50 MiB) compared to the A100 (40 MiB), leading NVIDIA to implement a sub-NUMA microarchitecture within the local NUMA node on H100. Additionally, the latency overhead for accessing a remote sub-NUMA node is minimal, approximately \subllchoppergpugap{}.

\begin{figure}
    \centering
    \begin{subfigure}{0.49\linewidth}
    
    \includegraphics[width=\linewidth]{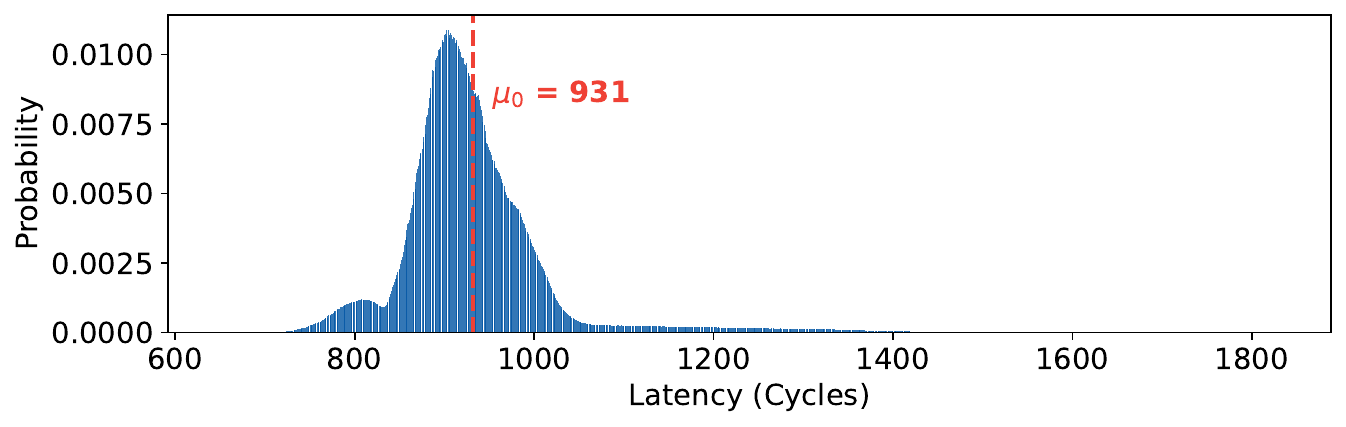}
    \caption{Original data on the RTX 5090.}
    \label{fig:dramlatency:allsm:rtx5090:raw}
    \end{subfigure}
    \begin{subfigure}{0.49\linewidth}
    \includegraphics[width=\linewidth]{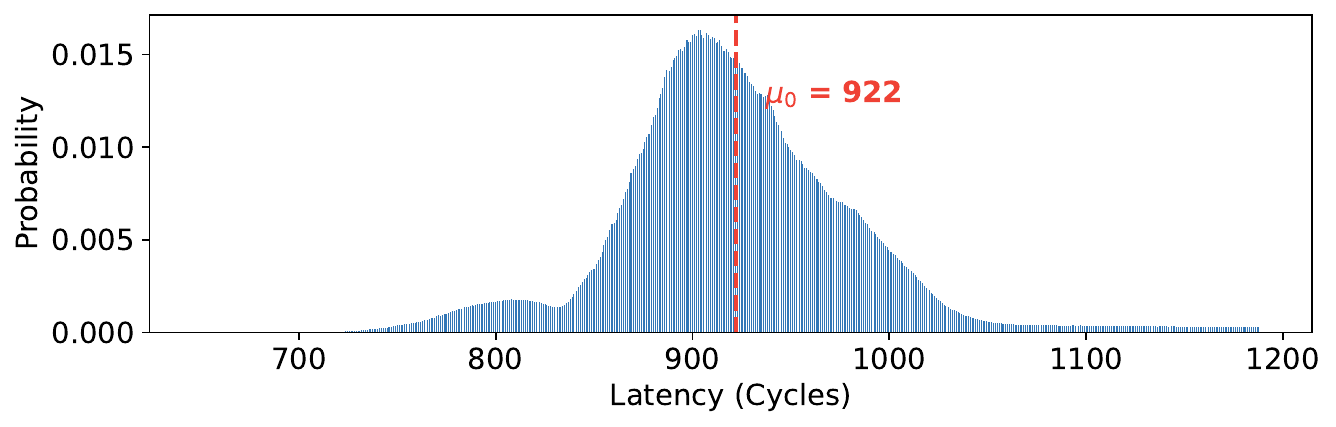}
    \caption{Data removed outliers on the RTX 5090 }
    \label{fig:dramlatency:allsm:rtx5090}
    \end{subfigure}
    \caption{\tacorebuttal{The distribution of access latencies for all SMs on GPUs accessing all DRAM NUMA nodes follows a Gaussian distribution for RTX 5090. These latencies are clusterred around a single central value: \dramblackgpu{} for the RTX 5090 after removing outliers.}}
    \label{fig:dramlatency:rtx5090:allsm}
\end{figure}

\noindent\tacorebuttal{\textbf{RTX 5090} We further apply \projname{} to the RTX 5090, which is a desktop GPU based on the Blackwell architecture and released in January 2025. As shown in \autoref{fig:dramlatency:rtx5090:allsm}, DGNA reveals that the RTX 5090 employs a single L2 cache partition, in contrast to the H100 and A100, which feature multiple L2 partitions. In addition, DGNA enables more accurate measurement of DRAM latencies by effectively filtering out outliers.} 

\subsection{NUMA Architecture Topology}
\label{sec:eval:numaarchtopology}
We next investigate the collected latency data to reveal the NUMA architecture and the correlations between DRAM and L2 NUMA nodes with the SMs on the A100 and H100.

\noindent\textbf{DRAM NUMA Architecture.}
We derive the relationship between GPU cores (SMs) and DRAM NUMA nodes on both A100 and H100 GPUs. Our analysis 
demonstrates two different patterns in relation to DRAM NUMA nodes.

\begin{figure}[t]
    \centering
    \begin{subfigure}{0.49\linewidth}
    \includegraphics[width=\linewidth]{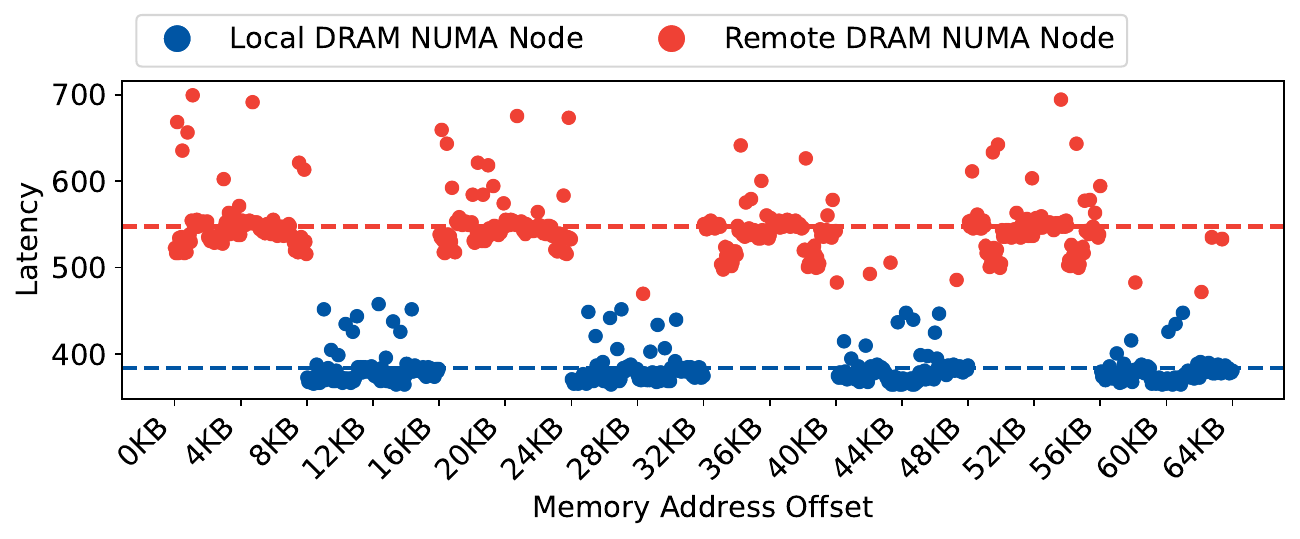}
      \caption{A100 $SM_0$}
    \label{fig:dramnuma:sm0:a100}
    \end{subfigure}
    \begin{subfigure}{0.49\linewidth}
     \includegraphics[width=\linewidth]{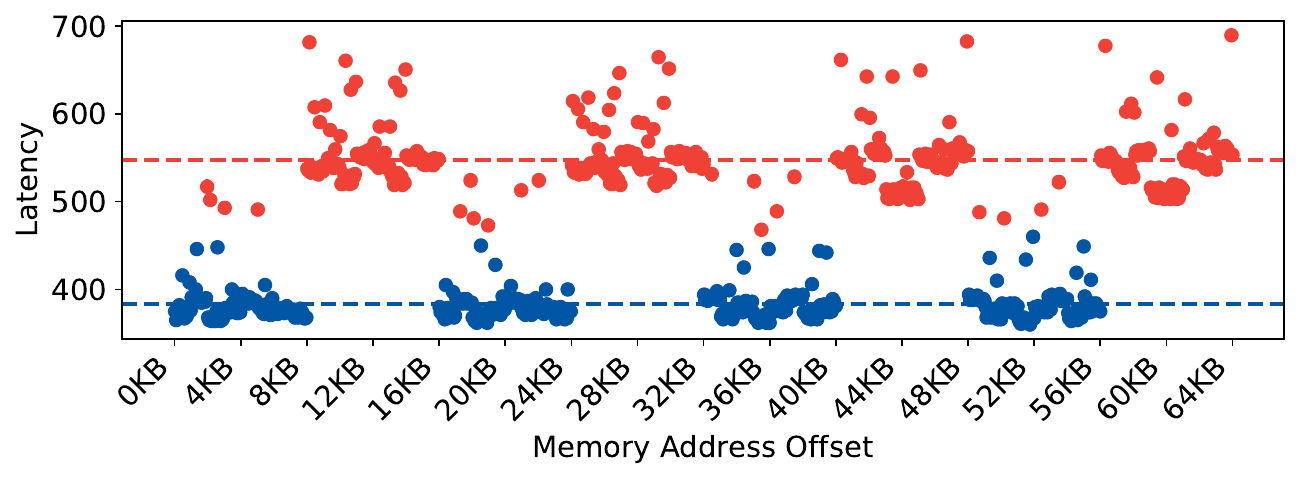}
    \caption{A100 $SM_2$}
    \label{fig:dramnuma:sm2:a100}
    \end{subfigure}
    \begin{subfigure}{0.49\linewidth}
    \includegraphics[width=\linewidth]{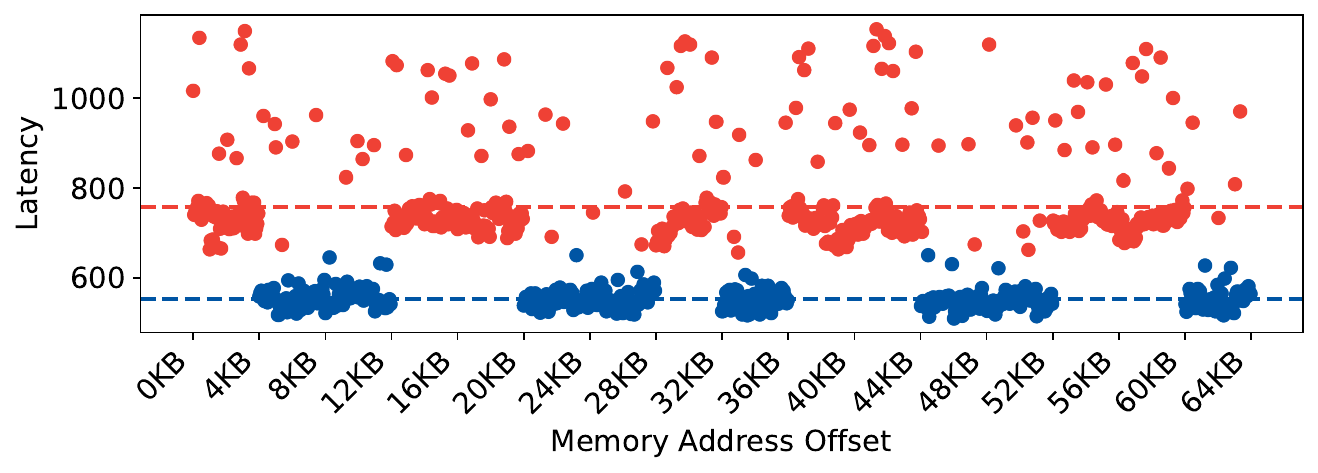}
     \caption{H100 $SM_0$}
    \label{fig:dramnuma:sm0:h100}
    \end{subfigure}
    \begin{subfigure}{0.49\linewidth}
     \includegraphics[width=\linewidth]{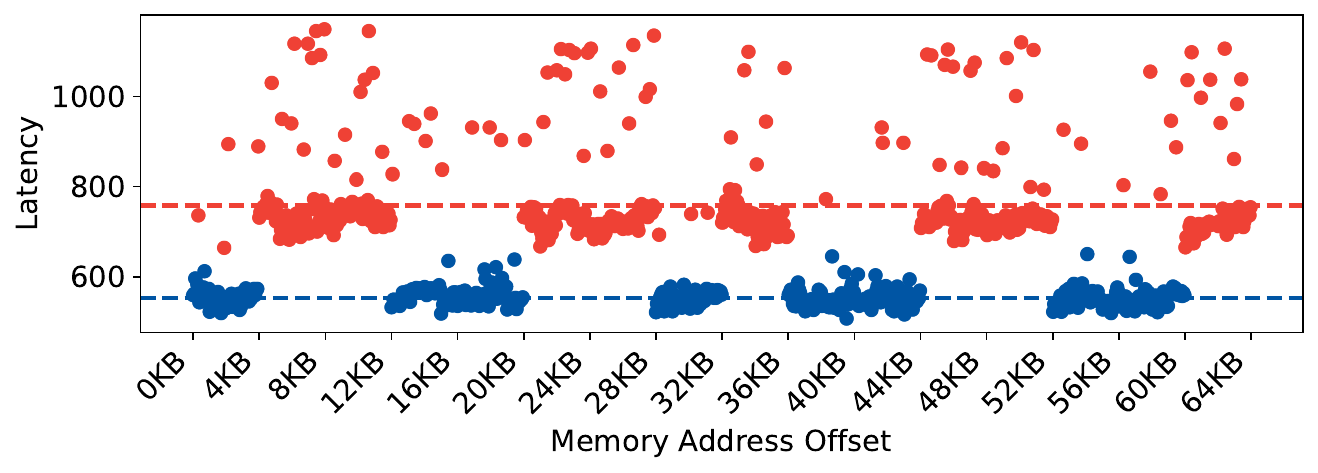}
    \caption{H100 $SM_2$}
    \label{fig:dramnuma:sm2:h100}
    \end{subfigure}
    \caption{We measure the latency feature vectors by accessing consecutive memory addresses through DRAM NUMA nodes. $SM_i$ indicates the SM whose ID is $i$.
    The dots are marked as blue if they are close to the average latency of the local DRAM NUMA node, and as red if they are close to the average latency of the remote DRAM NUMA node.}
    \label{fig:dramnuma}
\end{figure}

\tacorebuttal{\autoref{fig:dramnuma} demonstrates the NUMA allocation mechanisms on the A100 and H100. 
Each SMs on GPUs that access data from their local DRAM NUMA node exhibit lower memory latency values compared to those accessing data from a remote NUMA node. 
On the A100, the allocation alternates every 8KB, where each 8KB portion corresponds to a specific NUMA node. On the H100, The NUMA memory is allocated based on a formula that determines the allocation pattern by address offset. The formula is $ID_{NUMA}=\lfloor \frac{x+1-H(x-8)}{2} \rfloor mod\ 2$ where $x=\lfloor\frac{address\ mod\ 64KB}{4KB}\rfloor$ and H(x-8) is the Heaviside step function $H(x-8)=\begin{cases}1,&x\geqslant 8\\0,&x<8\end{cases}$. }

Furthermore, we group DRAM NUMA portions and the corresponding GPU cores using K-Means and the feature vectors. \autoref{tab:dramnuma:sms} demonstrates that on the A100, 46 SMs are assigned to DRAM NUMA node 0 and 62 SMs are assigned to DRAM NUMA node 1.
On the H100, both DRAM NUMA nodes have 66 corresponding SMs. 

This result demonstrates the differences between the A100 and H100 in handling die manufacturing errors. 
\tacorebuttal{NVIDIA employs the floorsweeping technique~\cite{bakita2023hardwarecomputepartitioning} to salvage imperfect dies by disabling defective hardware blocks.}
According to NVIDIA's official documentation~\cite{nvidia-a100,nvidia-h100}, the GA100 full GPU consists of 128 SMs, whereas the GH100 full GPU features 144 SMs. The A100 appears to combine two dies with different defective rates, while the H100 utilizes two dies with similar rates. This implies that NVIDIA improved the technique on the H100 to reduce the die defective rate compared to the A100.

\begin{table}
    \centering
    \caption{The corresponding relationship between GPU cores and NUMA memory partitions for A100 and H100. 
    We utilize $ID_{SM}$ to represent the SM ID. On the A100, we summarize the distribution of SMs using $\mathbf{X = ID_{SM} \mod 14}$. On the H100, we use $\mathbf{Y = ID_{SM} \mod 16} \quad \text{when} \quad \mathbf{ID_{SM} < 62}$, and $\mathbf{Z = (ID_{SM} - 62) \mod 14} \quad \text{when} \quad 62 \leq \mathbf{ID_{SM}} < 120$. The number of corresponding SMs for each NUMA node is 62 and 46 for the A100, which is imbalanced, whereas it is 66 and 66 for the H100, which is balanced.}
    \begin{tabular}{ccc}
        \toprule
        GPUs&DRAM NUMA Node 0 &DRAM NUMA Node 1\\
        \midrule
        \multirow{2}{*}{A100}&$0\leq X<2$ & $2\leq X<6$\\
        &$6\leq X<12$ &$12\leq X <14$ \\
        \midrule
         \multirow{6}{*}{H100} &$0\leq Y <2$  &$2\leq Y<8$ \\
         &$8\leq Y<14$  & $14\leq Y<16$  \\
         &$8\leq Z <14$  &$0\leq Z<8$ \\
         &$120\leq ID_{SM}<122$&$122\leq ID_{SM}<124$\\
         &$124\leq ID_{SM}<132$&\\
         \bottomrule
    \end{tabular}
    \label{tab:dramnuma:sms}
\end{table}

\begin{figure}[t]
    \centering
    \begin{subfigure}{0.49\linewidth}
    \includegraphics[width=\linewidth]{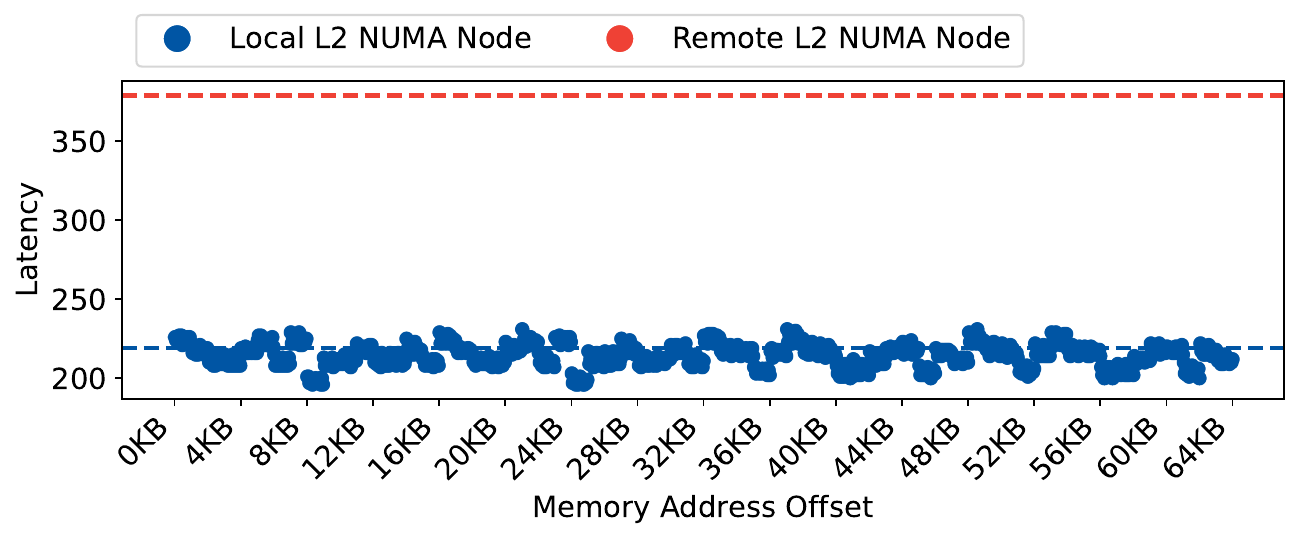}
      \caption{A100: using the SM on the same NUMA node to warmup.}
    \label{fig:l2numa:sm0:a100}
    \end{subfigure}
    \begin{subfigure}{0.49\linewidth}
     \includegraphics[width=\linewidth]{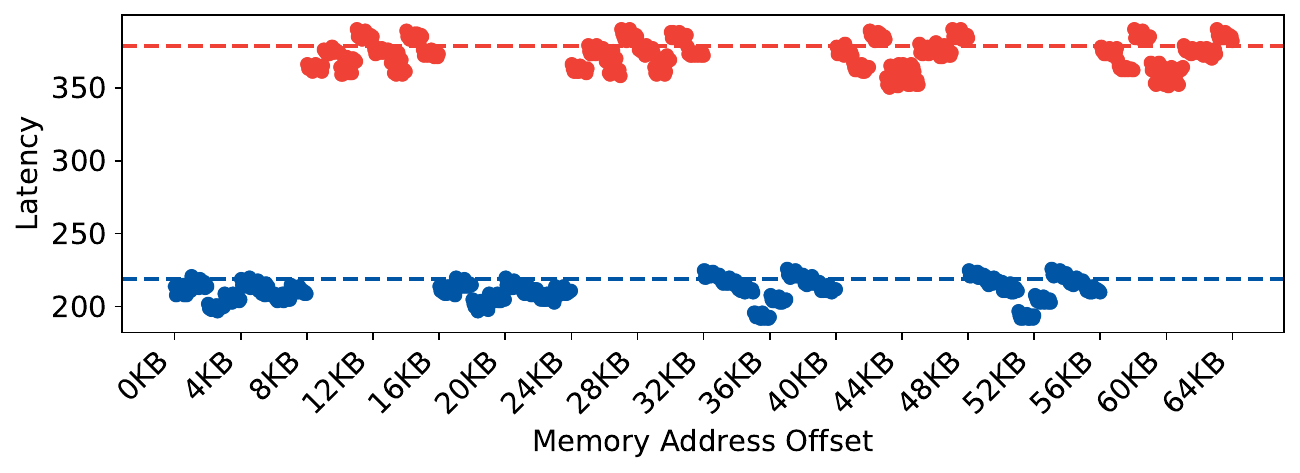}
    \caption{A100: using the SM on the different NUMA node to warmup.}
    \label{fig:l2numa:sm2:a100}
    \end{subfigure}
    \begin{subfigure}{0.49\linewidth}
    \includegraphics[width=\linewidth]{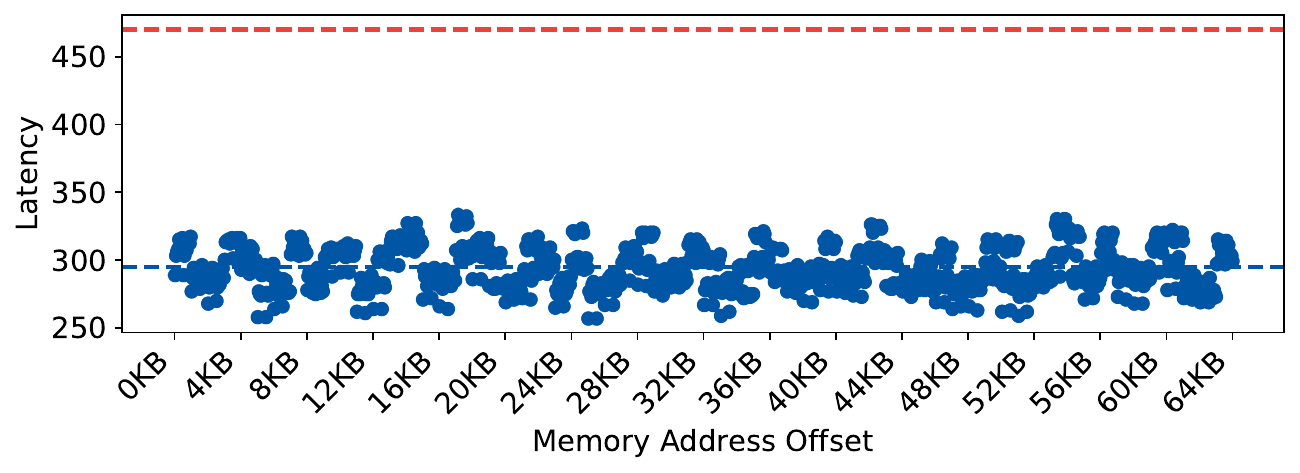}
     \caption{H100: using the SM on the same NUMA node to warmup.}
    \label{fig:l2numa:sm0:h100}
    \end{subfigure}
    \begin{subfigure}{0.49\linewidth}
     \includegraphics[width=\linewidth]{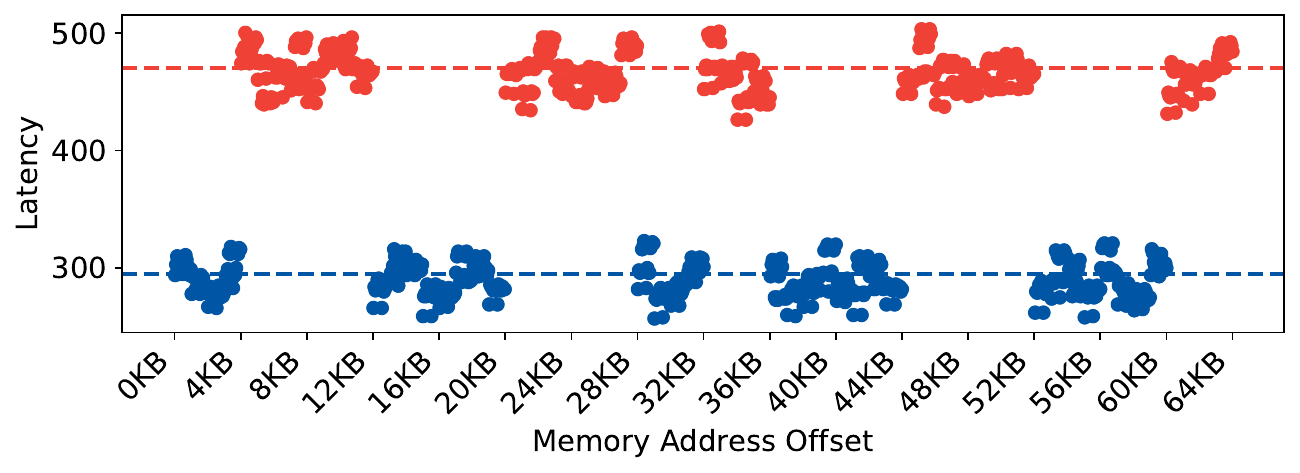}
    \caption{H100: using the SM on the different NUMA node to warmup. }
    \label{fig:l2numa:sm2:h100}
    \end{subfigure}
    \caption{Accessing consecutive memory addresses to derive the L2 NUMA node information. For measuring L2 latency within the same NUMA node, $SM_0$ fetches data into the L2 cache and then $SM_1$ is used to measure L2 latencies. To assess latency across different NUMA nodes, $SM_0$ fetches the data, but $SM_2$ is used to measure L2 latencies. The dots are marked as blue if they are close to the local L2 NUMA latency, and as red if they are close to the remote L2 NUMA latency.}
    \label{fig:l2numa}
\end{figure}

\noindent\textbf{L2 NUMA architecture.} We then utilize the feature vectors of L2 NUMA nodes to further cluster groups. However, we find that the results are identical to those obtained from clustering based on DRAM NUMA nodes.
We conclude that L2 and DRAM are directly connected, and the information transfer between them occurs through the connections between L2 NUMA nodes. As a result, L2 NUMA nodes cannot further split SMs into smaller groups.

Another piece of evidence supporting this conclusion is the latency gap between fetching data from the local L2 NUMA node versus a remote NUMA node, which is \dramamperegpugap{} and \dramhoppergpugap{} for the A100 and H100, respectively. Meanwhile, the overhead for SMs accessing remote L2 NUMA nodes is \llcamperegpugap{} for the A100 and \llchoppergpugap{} for the H100. This Network-on-Chip (NoC) overhead is similar to the L2 and DRAM latency for SMs accessing remote NUMA nodes, suggesting that the same wires are used for both.
\begin{figure}[t]
    \centering
    \begin{subfigure}{0.49\linewidth}
    \includegraphics[width=\linewidth]{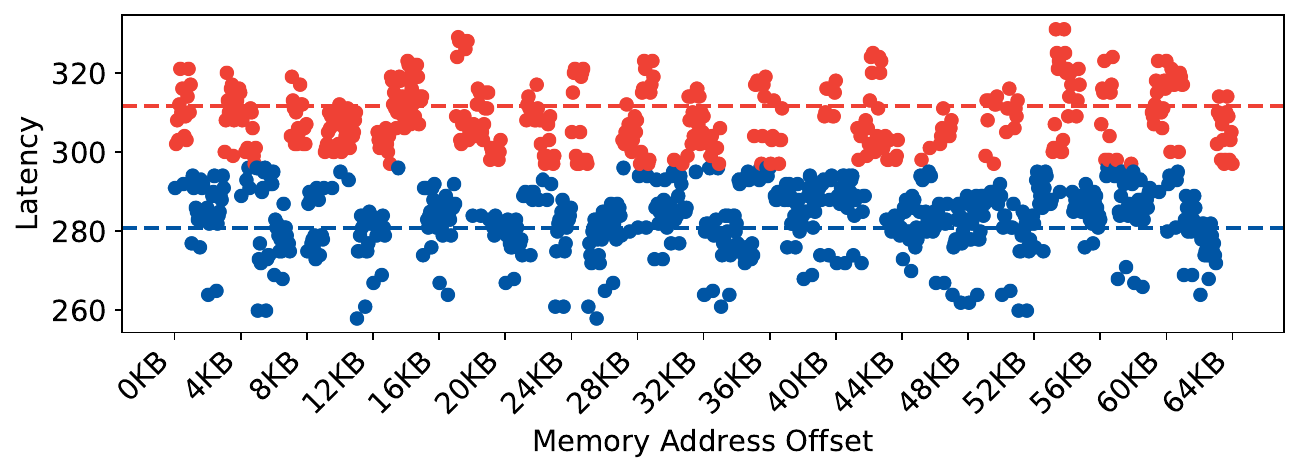}
     \caption{using the SM on the same sub-NUMA node to warmup.}
    \label{fig:subl2numa:sm0_1:h100}
    \end{subfigure}
    \begin{subfigure}{0.49\linewidth}
     \includegraphics[width=\linewidth]{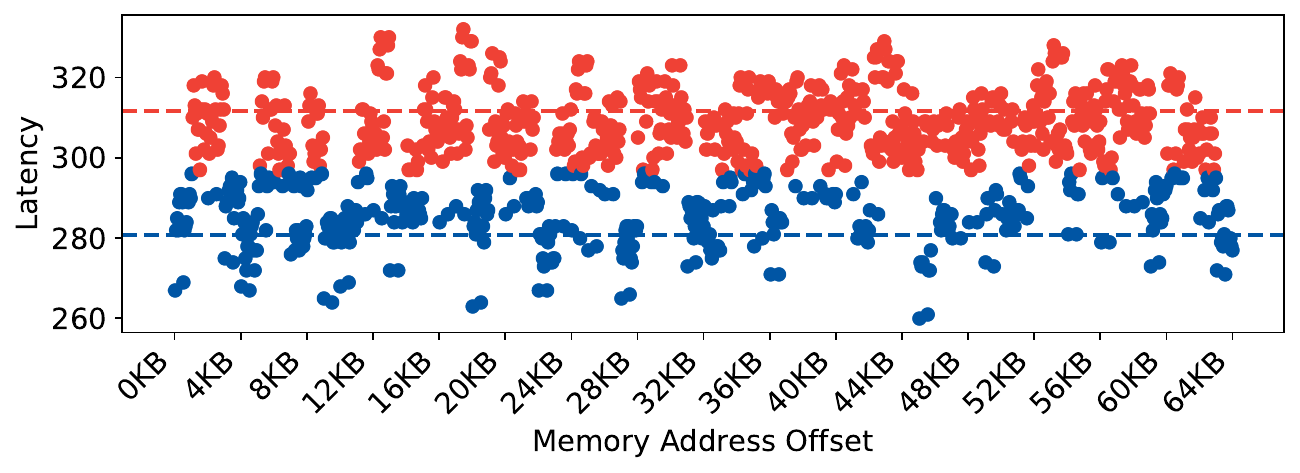}
    \caption{using the SM on the different sub-NUMA node to warmup.}
    \label{fig:subl2numa:sm0_8:h100}
    \end{subfigure}
    \caption{The sub-NUMA information is obtained by accessing consecutive memory addresses using different SMs to warm up and measure L2 latencies. Our findings reveal that the H100 contains sub-NUMA L2 nodes within each L2 local NUMA node. We mark the dots as blue if they are close to the average latency of the local sub-NUMA node, and as red if they are close to the average latency of the remote sub-NUMA node. The results indicate that the first 2KB of memory addresses exhibit high latency in (a) but low latency in (b), followed by alternating between these two patterns}
    \label{fig:subl2numa}
\end{figure}
\begin{figure}
    \centering
    \includegraphics[width=0.8\linewidth]{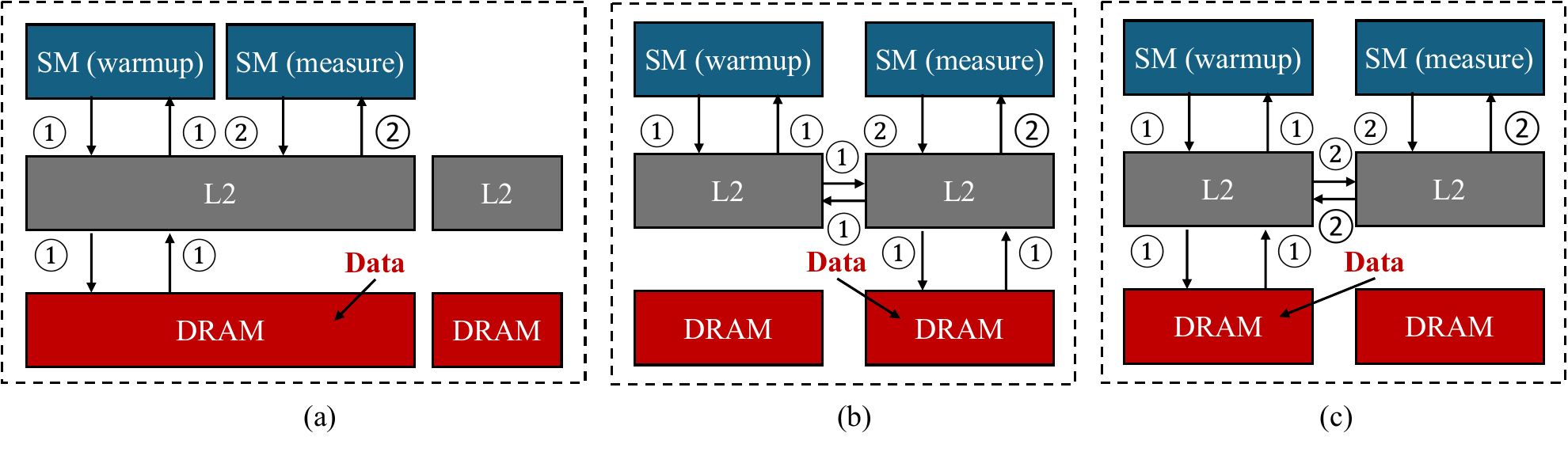}
    \caption{The mechanism of GPUs to read data on L2 NUMA includes 3 cases: \textbf{(a).} The home-SM warms up the local NUMA node, and another home-SM accesses it. \textbf{(b).} The remote-SM warms up both the local and remote NUMA nodes, and the home-SM accesses it. Remote-SM with remote-SM is similar to this case. \textbf{(c).} The home-SM warms up only the local NUMA node, and the remote-SM accesses it.}
    \label{fig:l2:machanism}
\end{figure}

\noindent\textbf{L2 sub-NUMA architecture.}
\autoref{fig:subl2numa} presents the results of using $SM_0$ to warm up the L2 cache, followed by measuring L2 access latency from both $SM_0$ and $SM_8$. Since the L1 cache is flushed after kernel execution, all these measurements reflect L2 latency.
Specifically, the latency to access the local sub-NUMA node is lower than that for accessing the remote sub-NUMA node, indicating that SM 0 and SM 8 are connected to different sub-NUMA nodes.

We further cluster SMs into four groups based on their relationship with sub-NUMA nodes, as shown in \autoref{tab:sub-SMS:distribute:h100}. 
Next, we use the thread block cluster on the H100 to analyze the SM IDs. Our analysis reveals that SMs connected to the same sub-NUMA node always belong to two GPU processing clusters (GPCs), indicating that two GPCs are connected to each sub-NUMA node on the H100. \autoref{tab:sub-SMS:distribute:h100} also illustrates that SMs within each L2 NUMA node are still imbalanced, 32 or 34 SMs per sub NUMA node. We conclude that this imbalance arises because GPCs are composed of texture processing clusters, each of which contains two SMs. As a result, no sub-NUMA node can be connected to exactly 33 SMs.

\begin{table}[t]
    \centering
    \caption{The corresponding relationship between SMs and NUMA/sub-NUMA nodes on the H100 reveals that SMs connected to a sub-NUMA node belong to one of the two GPCs. This indicates that GPCs are the fundamental units responsible for connecting to sub-NUMA nodes. 
    }
    \label{tab:sub-SMS:distribute:h100}
    \begin{tabular}{ccccc}
    \toprule
    NUMA&sub-NUMA&\#SM&GPC&SM ID\\
    \midrule
    \multirow{4}{*}{\rotatebox{90}{NUMA 0}}&\multirow{2}{*}{{sub-NUMA 0}}&\multirow{2}{*}{32}&$GPC_0$&\multicolumn{1}{c}{2, 3, 18, 19, 34, 35,  50, 51, 64, 65, 78, 79, 92, 93, 106, 107.}\\
    &&&\tacorebuttal{$GPC_1$}&\multicolumn{1}{c}{4, 5, 20, 21, 36, 37, 52, 53, 66, 67, 80, 81, 94, 95, 108, 109.}\\
    &\multirow{2}{*}{{sub-NUMA 1}}&\multirow{2}{*}{34}&$GPC_2$&\multicolumn{1}{c}{6, 7, 22, 23,  38, 39, 54, 55, 68, 69, 82, 83, 96, 97, 110, 111.}\\
    &&&$GPC_3$&\multicolumn{1}{c}{14, 15, 30, 31, 46, 47, 62, 63, 76, 77, 90, 91,  104, 105,  118, 119, 122, 123.}\\
    \multirow{4}{*}{\rotatebox{90}{NUMA 1 }}&\multirow{2}{*}{{sub-NUMA 2}}&\multirow{2}{*}{34}&$GPC_4$&\multicolumn{1}{c}{0, 1, 16, 17,  32, 33,  48, 49, 124, 125, 126, 127, 128, 129, 130, 131.}\\
    &&&$GPC_5$&\multicolumn{1}{c}{12, 13, 28, 29, 44, 45, 60, 61, 74, 75, 88, 89, 102, 103, 116, 117, 120, 121. }\\
    &\multirow{2}{*}{{sub-NUMA 3}}&\multirow{2}{*}{32}&$GPC_6$&\multicolumn{1}{c}{8, 9, 24, 25, 40, 41, 56, 57, 70, 71, 84, 85, 98, 99, 112, 113.}\\
    &&&$GPC_7$&\multicolumn{1}{c}{10, 11, 26, 27, 42, 43, 58, 59, 72, 73, 86, 87, 100, 101, 114, 115.}\\
    \bottomrule
    \end{tabular}
\end{table}

\subsection{Read and write mechanisms on the L2 NUMA}
\label{sec:eval:readwrite}
We then analyze the read and write mechanism on the L2 NUMA. 
Using the NUMA allocation results in \autoref{sec:numaarchtopology}, we can control the home, remote SM, L2 and DRAMs.

\noindent\textbf{Read operations.} As shown in \autoref{fig:l2numa}, Since $SM_2$ resides on the same L2 NUMA node as $SM_1$, their L2 access latencies are similar, both being close to \llcamperegpulow{} for the A100 and \llchoppergpulow{} for the H100.

\tacorebuttal{Moreover, we observe that L2 access latency varies across addresses when accessed by $SM_2$. Comparing these results with \autoref{fig:dramnuma}, we find that the L2 access latency matches that of a local L2 NUMA node when the SM warms up the L2 cache using data from a remote DRAM NUMA node. In contrast, when the data originates from the local DRAM NUMA node, the observed L2 access latency is closer to that of a remote L2 NUMA node.}


We conclude that the read mechanism on L2 NUMA of both the A100 and H100 operate as shown in \autoref{fig:l2:machanism}. There are 3 cases for L2 NUMA nodes reading data. (1). As shown in \autoref{fig:l2:machanism}a, the home-SM warms up only the local L2 NUMA node, and another home-SM accesses it, resulting in L2 access latency similar to that of accessing the local NUMA node. (2). As shown in \autoref{fig:l2:machanism}b, the home-SM warms up only the local NUMA node, and the remote-SM accesses it, resulting in L2 access latency similar to that of accessing the remote NUMA node. (3). As shown in \autoref{fig:l2:machanism}c, the remote-SM warms up both the local and remote NUMA nodes, the latency of both remote-SM and home-SM accessing data block is similar to that of accessing the local NUMA node.

We also notice that, on both the A100 or H100, the L1 cache is flushed after the kernel finishes executing, but the data inside the L2 NUMA node remains. When utilizing the same SM to warm up and measure data, the results are similar to those of accessing the local L2 NUMA node. 

\begin{table}[ht]
    \centering
    \caption{The results of Listing~\ref{listing:testwrite:code} measure the write mechanism of GPUs on L2 NUMA nodes. The warmup phase includes three different cases: warmup by home-SM, remote-SM, and both. We test writes by home-SM and remote-SM under after warmup. Subsequently, we measure the latency of read-after-write for both home-SM and remote-SM. The results reveal four patterns on both the A100 and H100, indicating that home-SM always writes to the local NUMA node if there is a cache hit. Remote-SM only updates the remote-NUMA if and only if the remote-NUMA cache hits, otherwise it updates all L2 NUMA nodes.}
    \begin{tabular}{ccccccc}
        \toprule
        Warmup&\multicolumn{2}{c}{Home-SM}&\multicolumn{2}{c}{Remote-SM}&\multicolumn{2}{c}{Both}\\
        Write&\multicolumn{1}{c}{Home-SM}&\multicolumn{1}{c}{Remote-SM}&  \multicolumn{1}{c}{Home SM}&\multicolumn{1}{c}{Remote-SM}&\multicolumn{1}{c}{Home-SM}&\multicolumn{1}{c}{Remote-SM}\\
        \midrule
        A100-Write&\textcolor{darkgreen}{281}&\textcolor{darkred}{425}&\textcolor{darkred}{385}&\textcolor{darkred}{407}&\textcolor{darkred}{411}&\textcolor{darkred}{446}\\
A100-RAW-Home-SM&\textcolor{darkgreen}{223}&\textcolor{darkgreen}{209}&\textcolor{darkgreen}{223}&\textcolor{darkred}{365}&\textcolor{darkgreen}{205}&\textcolor{darkgreen}{208}\\
A100-RAW-Remote-SM&\textcolor{darkred}{388}&\textcolor{darkgreen}{205}&\textcolor{darkred}{384}&\textcolor{darkgreen}{206}&\textcolor{darkred}{366}&\textcolor{darkgreen}{205}\\
\midrule
        H100-Write&\textcolor{darkgreen}{426}&\textcolor{darkred}{595}&\textcolor{darkred}{502}&\textcolor{darkred}{591}&\textcolor{darkred}{556}&\textcolor{darkred}{594}\\
        H100-RAW-Home-SM&\textcolor{darkgreen}{281}&\textcolor{darkgreen}{299}&\textcolor{darkgreen}{279}&\textcolor{darkred}{472}&\textcolor{darkgreen}{301}&\textcolor{darkgreen}{297}\\
        H100-RAW-Remote-SM&\textcolor{darkred}{444}&\textcolor{darkgreen}{295}&\textcolor{darkred}{447}&\textcolor{darkgreen}{295}&\textcolor{darkred}{473}&\textcolor{darkgreen}{298}\\
         \bottomrule        
    \end{tabular}
    \label{tab:writemachanism}
\end{table}

\textbf{Write operations.} 
The results, as shown in \autoref{tab:writemachanism}, present the write latency for different cases on the A100 and H100. These cases include: (a) whether the data is inside all L2 NUMA nodes or only in the local or remote L2 NUMA node, and (b) the location of the data inside DRAM, either in the SM's local DRAM NUMA node or in the remote DRAM NUMA node.
Additionally, \autoref{tab:writemachanism} presents the read latencies following a write, considering two different cases on the A100 and H100: (a) reading using the home SM, and (b) reading using the remote SM.


In the case of \textit{Home-SM (Warmup)-Home-SM (Write)}, the home SM fetches data only into the local NUMA node, resulting in the lowest write latency, as it only needs to update the local NUMA node. When the home SM reloads the data from the local NUMA node, the latency remains low for Home-SM (RAW). However, for Remote-SM (RAW), the remote NUMA node does not buffer this cache block, leading to a high latency for the remote SM.
For \textit{Home-SM (Warmup)-Remote-SM (Write)}, the remote SM updates all L2 NUMA nodes. Subsequently, when SMs reload the data from the L2 caches, the latency remains low for both Home-SM (RAW) and Remote-SM (RAW).

In the \textit{Remote-SM (Warmup)-Home-SM (Write)} case, the remote SM fetches data into both the remote and local NUMA nodes. 
The home SM updates only the local NUMA node, invalidating the remote NUMA node. Thus, Home-SM (RAW) latency remains low. For Remote-SM (RAW), its latency is high as the remote NUMA node does not buffer the latest cache block.
For \textit{Remote-SM (Warmup)-Remote-SM (Write)}, the remote SM updates only the remote L2 NUMA nodes. When the home SM reloads the data, it requires access to the remote L2 NUMA, resulting in high latency for Home-SM (RAW). However, Remote-SM (RAW) latency is low, as the L2 cache buffers the most recent data block.

For \textit{Both (Warmup)-Home-SM (Write)}, the write operation only updates home NUMA node. But for \textit{Both (Warmup)-Remote-SM (Write)}, both remote and home NUMA nodes are updated. The difference between \textit{Remote (Warmup)-Remote-SM (Write)} and \textit{Both (Warmup)-Remote-SM (Write)} is because when the remote SM first touches the data, the primary NUMA node for the data block is maintained in the remote SM's L2 node, despite also buffering data into the home L2 NUMA node~\cite{lenoski1990directorybasedcacheprotocol,franques2021widirdirectorybasedcacheprotocol}. 
However, in \textit{Both (Warmup)-Home-SM (Write)}, the primary NUMA node is the home L2. 

\section{Conclusion}
In this work, we present \projname{}, a framework to uncover the NUMA architecture within the GPU memory hierarchy through microbenchmarking and data analysis. 
\projname{} independently measures L2 and DRAM latencies without relying on vendor-specific instructions, ensuring accurate metrics by using a Gaussian mixture model to filter outliers. Our findings 
identify the presence of sub-NUMA nodes on the H100. Additionally, we uncover the read and write mechanism on the L2 NUMA nodes,
which employs different mechanisms across home and remote NUMA nodes for home and remote SMs. 
\projname{} will be open-sourced to support continued research.

\bibliographystyle{ACM-Reference-Format}
\bibliography{references}

@String{Computing = "Computing" }

@String{Computer = "{IEEE} Computer" }

@String{Springer = "Springer-Verlag" }

@inproceedings{nair2013casedramrefreshing,
  title={A case for refresh pausing in DRAM memory systems},
  author={Nair, Prashant and Chou, Chia-Chen and Qureshi, Moinuddin K},
  booktitle={International Symposium on High Performance Computer Architecture (HPCA)},
  _pages={627--638},
  year={2013},
  _organization={IEEE}
}

@inproceedings{dalmia2024cpelidegpucoherence,
  title={CPElide: Efficient Multi-Chiplet GPU Implicit Synchronization},
  author={Dalmia, Preyesh and Kumar, Rajesh Shashi and Sinclair, Matthew D},
  booktitle={International Symposium on Microarchitecture (MICRO)},
  _pages={700--717},
  year={2024},
  _organization={IEEE}
}

@inproceedings{alsop2016lazygpuconsistency,
  title={Lazy release consistency for GPUs},
  author={Alsop, Johnathan and Orr, Marc S and Beckmann, Bradford M and Wood, David A},
  booktitle={International Symposium on Microarchitecture (MICRO)},
  pages={1--14},
  year={2016},
  _organization={IEEE}
}

@inproceedings{ren2020hmggpucoherence,
  title={Hmg: Extending cache coherence protocols across modern hierarchical multi-gpu systems},
  author={Ren, Xiaowei and Lustig, Daniel and Bolotin, Evgeny and Jaleel, Aamer and Villa, Oreste and Nellans, David},
  booktitle={International Symposium on High Performance Computer Architecture (HPCA)},
  _pages={582--595},
  year={2020},
  _organization={IEEE}
}

@article{koukos2016buildinggpucoherence,
  title={Building heterogeneous unified virtual memories (uvms) without the overhead},
  author={Koukos, Konstantinos and Ros, Alberto and Hagersten, Erik and Kaxiras, Stefanos},
  journal={Transactions on Architecture and Code Optimization (TACO)},
  _volume={13},
  _number={1},
  _pages={1--22},
  year={2016},
  _publisher={ACM New York, NY, USA}
}

@inproceedings{alsop2018spandexgpucoherence,
  title={Spandex: A flexible interface for efficient heterogeneous coherence},
  author={Alsop, Johnathan and Sinclair, Matthew and Adve, Sarita},
  booktitle={International Symposium on Computer Architecture (ISCA)},
  _pages={261--274},
  year={2018},
  _organization={IEEE}
}

@inproceedings{young2018combininggpunuma,
  title={Combining HW/SW mechanisms to improve NUMA performance of multi-GPU systems},
  author={Young, Vinson and Jaleel, Aamer and Bolotin, Evgeny and Ebrahimi, Eiman and Nellans, David and Villa, Oreste},
  booktitle={International Symposium on Microarchitecture (MICRO)},
  _pages={339--351},
  year={2018},
  _organization={IEEE}
}

@inproceedings{arunkumar2017mcmgpunuma,
  title={MCM-GPU: Multi-chip-module GPUs for continued performance scalability},
  author={Arunkumar, Akhil and Bolotin, Evgeny and Cho, Benjamin and Milic, Ugljesa and Ebrahimi, Eiman and Villa, Oreste and Jaleel, Aamer and Wu, Carole-Jean and Nellans, David},
  booktitle={International Symposium on Computer Architecture (ISCA)},
  _pages={320--332},
  year={2017},
  _organization={IEEE Computer Society}
}

@inproceedings{bakita2023hardwarecomputepartitioning,
  title={Hardware compute partitioning on NVIDIA GPUs},
  author={Bakita, Joshua and Anderson, James H},
  booktitle={Real-Time and Embedded Technology and Applications Symposium (RTAS)},
  _pages={54--66},
  year={2023},
  _organization={IEEE}
}

@inproceedings{jin2024uncoveringgpunoc,
  title={Uncovering Real GPU NoC Characteristics: Implications on Interconnect Architecture},
  author={Jin, Zhixian and Rocca, Christopher and Kim, Jiho and Kasan, Hans and Rhu, Minsoo and Bakhoda, Ali and Aamodt, Tor M and Kim, John},
  booktitle={International Symposium on Microarchitecture (MICRO)},
  _pages={885--898},
  year={2024},
  _organization={IEEE}
}

@inproceedings{franques2021widirdirectorybasedcacheprotocol,
  title={Widir: A wireless-enabled directory cache coherence protocol},
  author={Franques, Antonio and Kokolis, Apostolos and Abadal, Sergi and Fernando, Vimuth and Misailovic, Sasa and Torrellas, Josep},
  booktitle={International Symposium on High-Performance Computer Architecture (HPCA)},
  _pages={304--317},
  year={2021},
  _organization={IEEE}
}

@inproceedings{zhao2023nuba,
  title={NUBA: Non-uniform bandwidth GPUs},
  author={Zhao, Xia and Jahre, Magnus and Tang, Yuhua and Zhang, Guangda and Eeckhout, Lieven},
  booktitle={International Conference on Architectural Support for Programming Languages and Operating Systems (ASPLOS)},
  _pages={544--559},
  year={2023}
}

@inproceedings{lenoski1990directorybasedcacheprotocol,
  title={The directory-based cache coherence protocol for the DASH multiprocessor},
  author={Lenoski, Daniel and Laudon, James and Gharachorloo, Kourosh and Gupta, Anoop and Hennessy, John},
  booktitle={International symposium on Computer Architecture (ISCA)},
  _pages={148--159},
  year={1990}
}

@inproceedings{alappat2020understandingsubnuma,
  title={Understanding HPC benchmark performance on Intel Broadwell and Cascade Lake processors},
  author={Alappat, Christie L and Hofmann, Johannes and Hager, Georg and Fehske, Holger and Bishop, Alan R and Wellein, Gerhard},
  booktitle={International Conference on High Performance Computing (ISC)},
  _pages={412--433},
  year={2020},
  _organization={Springer}
}

@inproceedings{mukundan2013understandingrefreshing,
  title={Understanding and mitigating refresh overheads in high-density DDR4 DRAM systems},
  author={Mukundan, Janani and Hunter, Hillery and Kim, Kyu-Hyoun and Stuecheli, Jeffrey and Mart{\'\i}nez, Jos{\'e} F},
  booktitle={International Symposium on Computer Architecture (ISCA)},
  year={2013}
}

@article{jia2018dissectingvoltagpu,
  title={Dissecting the NVIDIA volta GPU architecture via microbenchmarking},
  author={Jia, Zhe and Maggioni, Marco and Staiger, Benjamin and Scarpazza, Daniele P},
  journal={arXiv preprint arXiv:1804.06826},
  year={2018}
}

@article{jia2019dissectingturing,
  title={Dissecting the nvidia turing t4 gpu via microbenchmarking},
  author={Jia, Zhe and Maggioni, Marco and Smith, Jeffrey and Scarpazza, Daniele Paolo},
  journal={arXiv preprint arXiv:1903.07486},
  year={2019}
}

@article{sun2023dissectingtensorcores,
  title={Dissecting Tensor Cores via Microbenchmarks: Latency, Throughput and Numeric Behaviors},
  author={Sun, Wei and Li, Ang and Geng, Tong and Stuijk, Sander and Corporaal, Henk},
  journal={Transactions on Parallel \& Distributed Systems (TPDS)},
  _volume={34},
  _number={01},
  _pages={246--261},
  year={2023},
  _publisher={IEEE Computer Society}
}

@inproceedings{agarwal2015unlockinggpunuma,
  title={Unlocking bandwidth for GPUs in CC-NUMA systems},
  author={Agarwal, Neha and Nellans, David and O'Connor, Mike and Keckler, Stephen W and Wenisch, Thomas F},
  booktitle={International Symposium on High Performance Computer Architecture (HPCA)},
  _pages={354--365},
  year={2015},
  _organization={IEEE}
}

@inproceedings{thomas2020optimizingamperegpu,
  title={Optimizing cuda applications for nvidia a100 gpu},
  author={Thomas-Collignon, G and Mehta, V},
  booktitle={NVIDIA GPU Technology Conference},
  year={2020}
}

@inproceedings{ren2020hmggpunuma,
  title={Hmg: Extending cache coherence protocols across modern hierarchical multi-gpu systems},
  author={Ren, Xiaowei and Lustig, Daniel and Bolotin, Evgeny and Jaleel, Aamer and Villa, Oreste and Nellans, David},
  booktitle={International Symposium on High Performance Computer Architecture (HPCA)},
  _pages={582--595},
  year={2020},
  _organization={IEEE}
}

@inproceedings{DBLP:conf/isca/XieFCS19gpunuma,
  author       = {Chenhao Xie and
                  Xin Fu and
                  Mingsong Chen and
                  Shuaiwen Leon Song},
  title        = {{OO-VR:} {NUMA} friendly object-oriented {VR} rendering framework
                  for future NUMA-based multi-GPU systems},
  booktitle    = {International Symposium on Computer Architecture (ISCA)},
  _pages        = {53--65},
  _publisher    = {{ACM}},
  year         = {2019},
}

@inproceedings{li2023transgpunuma,
  title={Trans-fw: Short circuiting page table walk in multi-gpu systems via remote forwarding},
  author={Li, Bingyao and Yin, Jieming and Holey, Anup and Zhang, Youtao and Yang, Jun and Tang, Xulong},
  booktitle={International Symposium on High-Performance Computer Architecture (HPCA)},
  _pages={456--470},
  year={2023},
  _organization={IEEE}
}

@inproceedings{li2023orchestratedgpunuma,
  title={Orchestrated scheduling and partitioning for improved address translation in gpus},
  author={Li, Bingyao and Wang, Yueqi and Tang, Xulong},
  booktitle={Design Automation Conference (DAC)},
  pages={1--6},
  year={2023},
  organization={IEEE}
}

@inproceedings{lee2023snakebytegpunuma,
  title={Snakebyte: A tlb design with adaptive and recursive page merging in gpus},
  author={Lee, Jiwon and Lee, Ju Min and Oh, Yunho and Song, William J and Ro, Won Woo},
  booktitle={International Symposium on High-Performance Computer Architecture (HPCA)},
  _pages={1195--1207},
  year={2023},
  _organization={IEEE}
}

@inproceedings{milic2017beyondgpunuma,
  title={Beyond the socket: NUMA-aware GPUs},
  author={Milic, Ugljesa and Villa, Oreste and Bolotin, Evgeny and Arunkumar, Akhil and Ebrahimi, Eiman and Jaleel, Aamer and Ramirez, Alex and Nellans, David},
  booktitle={International Symposium on Microarchitecture (MICRO)},
  _pages={123--135},
  year={2017}
}

@inproceedings{wang2024gritgpunuma,
  title={GRIT: Enhancing Multi-GPU Performance with Fine-Grained Dynamic Page Placement},
  author={Wang, Yueqi and Li, Bingyao and Jaleel, Aamer and Yang, Jun and Tang, Xulong},
  booktitle={International Symposium on High-Performance Computer Architecture (HPCA)},
  _pages={1080--1094},
  year={2024},
  _organization={IEEE}
}

@inproceedings{prakash2016improvinggpugaming,
  title={Improving mobile gaming performance through cooperative CPU-GPU thermal management},
  author={Prakash, Alok and Amrouch, Hussam and Shafique, Muhammad and Mitra, Tulika and Henkel, J{\"o}rg},
  booktitle={Proceedings of the annual design automation conference (DAC)},
  _pages={1--6},
  year={2016}
}

@inproceedings{park2020blackmirrorgpugaming,
  title={Blackmirror: Preventing wallhacks in 3d online fps games},
  author={Park, Seonghyun and Ahmad, Adil and Lee, Byoungyoung},
  booktitle={ACM Conference on Computer and Communications Security (CCS)},
  pages={987--1000},
  year={2020}
}

@inproceedings{gu2024gvulkangpuraytracing,
  title={$\{$gVulkan$\}$: Scalable $\{$GPU$\}$ Pooling for $\{$Pixel-Grained$\}$ Rendering in Ray Tracing},
  author={Gu, Yicheng and Wang, Yun and Sun, Yunfan and Xiang, Yuxin and Hu, Xuyan and Qi, Zhengwei and Guan, Haibing},
  booktitle={USENIX Annual Technical Conference (USENIX ATC)},
  _pages={1151--1165},
  year={2024}
}

@inproceedings{ren2021chopingpugraphicsrendering,
  title={Chopin: Scalable graphics rendering in multi-gpu systems via parallel image composition},
  author={Ren, Xiaowei and Lis, Mieszko},
  booktitle={International Symposium on High-Performance Computer Architecture (HPCA)},
  _pages={709--722},
  year={2021},
  _organization={IEEE}
}

@inproceedings{dissectinghopperarchitecture,
  author       = {Weile Luo and
                  Ruibo Fan and
                  Zeyu Li and
                  Dayou Du and
                  Qiang Wang and
                  Xiaowen Chu},
  title        = {Benchmarking and Dissecting the Nvidia Hopper {GPU} Architecture},
  booktitle    = {International Parallel and Distributed Processing Symposium (IPDPS)},
  _pages        = {656--667},
  _publisher    = {{IEEE}},
  year         = {2024},
  _url          = {https://doi.org/10.1109/IPDPS57955.2024.00064},
  _doi          = {10.1109/IPDPS57955.2024.00064},
  _timestamp    = {Wed, 17 Jul 2024 15:59:37 +0200},
  _biburl       = {https://dblp.org/rec/conf/ipps/LuoFLDWC24.bib},
  _bibsource    = {dblp computer science bibliography, https://dblp.org}
}

@inproceedings{dutta2023spyinthegpubox,
  title={Spy in the GPU-box: Covert and side channel attacks on multi-GPU systems},
  author={Dutta, Sankha Baran and Naghibijouybari, Hoda and Gupta, Arjun and Abu-Ghazaleh, Nael and Marquez, Andres and Barker, Kevin},
  booktitle={International Symposium on Computer Architecture (ISCA)},
  _pages={1--13},
  year={2023}
}

@article{achiam2023gpt,
  title={Gpt-4 technical report},
  author={OpenAI},
  journal={arXiv preprint arXiv:2303.08774},
  year={2023}
}

@article{llamallm,
  author       = {Hugo Touvron and
                  Thibaut Lavril and
                  Gautier Izacard and
                  Xavier Martinet and
                  Marie{-}Anne Lachaux and
                  Timoth{\'{e}}e Lacroix and
                  Baptiste Rozi{\`{e}}re and
                  Naman Goyal and
                  Eric Hambro and
                  Faisal Azhar and
                  Aur{\'{e}}lien Rodriguez and
                  Armand Joulin and
                  Edouard Grave and
                  Guillaume Lample},
  title        = {LLaMA: Open and Efficient Foundation Language Models},
  journal      = {CoRR},
  _volume       = {abs/2302.13971},
  year         = {2023},
  _url          = {https://doi.org/10.48550/arXiv.2302.13971},
  _doi          = {10.48550/ARXIV.2302.13971},
  eprinttype    = {arXiv},
  eprint       = {2302.13971},
  _timestamp    = {Mon, 28 Aug 2023 21:26:20 +0200},
  biburl       = {https://dblp.org/rec/journals/corr/abs-2302-13971.bib},
  _bibsource    = {dblp computer science bibliography, https://dblp.org}
}

@inproceedings{zhang2015numagpunumaperf,
  title={NUMA-aware graph-structured analytics},
  author={Zhang, Kaiyuan and Chen, Rong and Chen, Haibo},
  booktitle={Symposium on principles and practice of parallel programming (PPoPP)},
  _pages={183--193},
  year={2015}
}

@article{roy2018numagpunumaperf,
  title={Numa-caffe: Numa-aware deep learning neural networks},
  author={Roy, Probir and Song, Shuaiwen Leon and Krishnamoorthy, Sriram and Vishnu, Abhinav and Sengupta, Dipanjan and Liu, Xu},
  journal={Transactions on Architecture and Code Optimization (TACO)},
  _volume={15},
  _number={2},
  _pages={1--26},
  year={2018},
  _publisher={ACM New York, NY, USA}
}

@inproceedings{xie2019oogpunumaperf,
  title={OO-VR: NUMA friendly object-oriented VR rendering framework for future NUMA-based multi-GPU systems},
  author={Xie, Chenhao and Xin, Fu and Chen, Mingsong and Song, Shuaiwen Leon},
  booktitle={International Symposium on Computer Architecture (ISCA)},
  _pages={53--65},
  year={2019},
  _organization={IEEE}
}

@inproceedings{zhang2024invalidategpusecurity,
  title={$\{$Invalidate+ Compare$\}$: A $\{$Timer-Free$\}$$\{$GPU$\}$ Cache Attack Primitive},
  author={Zhang, Zhenkai and Cai, Kunbei and Guo, Yanan and Yao, Fan and Gao, Xing},
  booktitle={USENIX Security Symposium (USENIX Security)},
  _pages={2101--2118},
  year={2024}
}

@inproceedings{damianou2012manifoldmultimodalgaussian,
  title={Manifold relevance determination},
  author={Damianou, AC and Ek, Carl Henrik and Titsias, MK and Lawrence, ND},
  booktitle={International Conference on Machine Learning (ICML)},
  _pages={145--152},
  year={2012}
}

@inproceedings{song2017multimodalgaussain,
  title={Multimodal Gaussian process latent variable models with harmonization},
  author={Song, Guoli and Wang, Shuhui and Huang, Qingming and Tian, Qi},
  booktitle={International Conference on Computer Vision (ICCV)},
  _pages={5029--5037},
  year={2017}
}

@article{ahmed2020kkmeans,
  title={The k-means algorithm: A comprehensive survey and performance evaluation},
  author={Ahmed, Mohiuddin and Seraj, Raihan and Islam, Syed Mohammed Shamsul},
  journal={Electronics},
  _volume={9},
  _number={8},
  _pages={1295},
  year={2020},
  _publisher={MDPI}
}

@inproceedings{dashti2013trafficnumahomenode,
  title={Traffic Management: A Holistic Approach to Memory Placement on NUMA Systems},
  author={Dashti, Mohammad and Fedorova, Alexandra and Funston, Justin and Gaud, Fabien and Lachaize, Renaud and Quema, Vivien and Roth, Mark},
  booktitle={International Conference on Architectural Support for Programming Languages and Operating Systems (ASPLOS)},
  year={2013}
}

@inproceedings{bera2022hermes,
  title={Hermes: Accelerating long-latency load requests via perceptron-based off-chip load prediction},
  author={Bera, Rahul and Kanellopoulos, Konstantinos and Balachandran, Shankar and Novo, David and Olgun, Ataberk and Sadrosadat, Mohammad and Mutlu, Onur},
  booktitle={International Symposium on Microarchitecture (MICRO)},
  pages={1--18},
  year={2022},
  _organization={IEEE}
}

@inproceedings{pakalapati2020bouquet,
  title={Bouquet of instruction pointers: Instruction pointer classifier-based spatial hardware prefetching},
  author={Pakalapati, Samuel and Panda, Biswabandan},
  booktitle={International Symposium on Computer Architecture (ISCA)},
  _pages={118--131},
  year={2020},
  _organization={IEEE}
}

@inproceedings{khairy2020accel,
  title={Accel-Sim: An extensible simulation framework for validated GPU modeling},
  author={Khairy, Mahmoud and Shen, Zhesheng and Aamodt, Tor M and Rogers, Timothy G},
  booktitle={ACM/IEEE 47th Annual International Symposium on Computer Architecture (ISCA)},
  pages={473--486},
  year={2020},
  _organization={IEEE}
}

@inproceedings{sun2019mgpusim,
  title={MGPUSim: enabling multi-GPU performance modeling and optimization},
  author       = {Yifan Sun and
                  Trinayan Baruah and
                  Saiful A. Mojumder and
                  Shi Dong and
                  Xiang Gong and
                  Shane Treadway and
                  Yuhui Bao and
                  Spencer Hance and
                  Carter McCardwell and
                  Vincent Zhao and
                  Harrison Barclay and
                  Amir Kavyan Ziabari and
                  Zhongliang Chen and
                  Rafael Ubal and
                  Jos{\'{e}} L. Abell{\'{a}}n and
                  John Kim and
                  Ajay Joshi and
                  David R. Kaeli},
  booktitle={Proceedings of the 46th International Symposium on Computer Architecture (ISCA)},
  pages={197--209},
  year={2019}
}

@inproceedings{chen2018tvm,
  author       = {Tianqi Chen and
                  Thierry Moreau and
                  Ziheng Jiang and
                  Lianmin Zheng and
                  Eddie Q. Yan and
                  Haichen Shen and
                  Meghan Cowan and
                  Leyuan Wang and
                  Yuwei Hu and
                  Luis Ceze and
                  Carlos Guestrin and
                  Arvind Krishnamurthy},
  title        = {{TVM:} An Automated End-to-End Optimizing Compiler for Deep Learning},
  booktitle    = {{USENIX} Symposium on Operating Systems Design and Implementation (OSDI)},
  pages        = {578--594},
  year         = {2018},
}

@inproceedings{abadi2016tensorflow,
  author       = {Mart{\'{\i}}n Abadi and
                  Paul Barham and
                  Jianmin Chen and
                  Zhifeng Chen and
                  Andy Davis and
                  Jeffrey Dean and
                  Matthieu Devin and
                  Sanjay Ghemawat and
                  Geoffrey Irving and
                  Michael Isard and
                  Manjunath Kudlur and
                  Josh Levenberg and
                  Rajat Monga and
                  Sherry Moore and
                  Derek Gordon Murray and
                  Benoit Steiner and
                  Paul A. Tucker and
                  Vijay Vasudevan and
                  Pete Warden and
                  Martin Wicke and
                  Yuan Yu and
                  Xiaoqiang Zheng},
  title        = {TensorFlow: {A} System for Large-Scale Machine Learning},
  booktitle    = {{USENIX} Symposium on Operating Systems Design and Implementation (OSDI)},
  pages        = {265--283},
  year         = {2016},
  _month = {November},
}

@misc{nvidia-h100,
  author = {NVIDIA},
  title = {{NVIDIA H100 Tensor Core GPU Architecture}},
  howpublished = {\url{https://resources.nvidia.com/en-us-tensor-core}},
  year = {2023},
key={},
}

@misc{nvidia-a100,
  author = {NVIDIA},
  title = {{NVIDIA A100 Tensor Core GPU Architecture}},
  howpublished = {\url{https://images.nvidia.com/aem-dam/en-zz/Solutions/data-center/nvidia-ampere-architecture-whitepaper.pdf}},
  year = {2020},
key={},
}

@inproceedings{alshboul2018lazypersistency,
  title={Lazy persistency: A high-performing and write-efficient software persistency technique},
  author={Alshboul, Mohammad and Tuck, James and Solihin, Yan},
  booktitle={International Symposium on Computer Architecture (ISCA)},
  pages={439--451},
  year={2018},
  _organization={IEEE}
}

@inproceedings{pytorch,
  author       = {Adam Paszke and
                  Sam Gross and
                  Francisco Massa and
                  Adam Lerer and
                  James Bradbury and
                  Gregory Chanan and
                  Trevor Killeen and
                  Zeming Lin and
                  Natalia Gimelshein and
                  Luca Antiga and
                  Alban Desmaison and
                  Andreas K{\"{o}}pf and
                  Edward Z. Yang and
                  Zachary DeVito and
                  Martin Raison and
                  Alykhan Tejani and
                  Sasank Chilamkurthy and
                  Benoit Steiner and
                  Lu Fang and
                  Junjie Bai and
                  Soumith Chintala},
 
  title        = {PyTorch: An Imperative Style, High-Performance Deep Learning Library},
  booktitle    = {Conference
                  on Neural Information Processing Systems (NeurIPS)},
  pages        = {8024--8035},
  year         = {2019},
}

@inproceedings{choquette2022nvidia,
  title={Nvidia hopper gpu: Scaling performance},
  author={Choquette, Jack},
  booktitle={Hot Chips 34 Symposium (HCS)},
  _pages={1--46},
  year={2022},
  _organization={IEEE Computer Society}
}

@inproceedings{sheng2023flexgengpullm,
  title={Flexgen: High-throughput generative inference of large language models with a single gpu},
  author={Sheng, Ying and Zheng, Lianmin and Yuan, Binhang and Li, Zhuohan and Ryabinin, Max and Chen, Beidi and Liang, Percy and R{\'e}, Christopher and Stoica, Ion and Zhang, Ce},
  booktitle={International Conference on Machine Learning (ICML)},
  _pages={31094--31116},
  year={2023},
  _organization={PMLR}
}

\end{document}